\documentclass{emulateapj}
\usepackage{natbib}
\usepackage[caption=false]{subfig}

\begin{document}
\title{MagAO Imaging of Long-period Objects (MILO). I. A Benchmark M Dwarf Companion Exciting a Massive Planet around the Sun-like Star HD 7449\footnotemark[*]}
\footnotetext[*]{This paper includes data obtained at the 6.5 m Magellan Telescopes located at Las Campanas Observatory, Chile.}
\author{Timothy J. Rodigas\altaffilmark{1,12}, Pamela Arriagada\altaffilmark{1}, Jackie Faherty\altaffilmark{1}, Guillem Anglada-Escud{\'e}\altaffilmark{2}, Nathan Kaib\altaffilmark{3}, R. Paul Butler\altaffilmark{1}, Stephen Shectman\altaffilmark{4}, Alycia Weinberger\altaffilmark{1}, Jared R. Males\altaffilmark{5,13}, Katie M. Morzinski\altaffilmark{5}, Laird M. Close\altaffilmark{5}, Philip M. Hinz\altaffilmark{5}, Jeffrey D. Crane\altaffilmark{4}, Ian Thompson\altaffilmark{4}, Johanna Teske\altaffilmark{1}, Mat\'ias D\'iaz\altaffilmark{4,6}, Dante Minniti\altaffilmark{7,8,9}, Mercedes Lopez-Morales\altaffilmark{10}, Fred C. Adams\altaffilmark{11}, Alan P. Boss\altaffilmark{1}}

\altaffiltext{1}{Department of Terrestrial Magnetism, Carnegie Institute of Washington, 5241 Broad Branch Road, NW, Washington, DC 20015, USA; email: trodigas@carnegiescience.edu}
\altaffiltext{2}{School of Physics and Astronomy, Queen Mary, University of London, 327 Mile End Rd. London, UK}
\altaffiltext{3}{Homer L. Dodge Department of Physics and Astronomy, The University of Oklahoma, 440 W. Brooks St. Norman, OK 73019, USA}
\altaffiltext{4}{The Observatories of the Carnegie Institution of Washington, 813 Santa Barbara Street, Pasadena, CA 91101, USA}
\altaffiltext{5}{Steward Observatory, The University of Arizona, 933 N. Cherry Ave., Tucson, AZ 85721, USA}
\altaffiltext{6}{Universidad de Chile, Departamento de Astronom\'ia, Camino El Observatorio 1515, Las Condes, Santiago, Chile.}
\altaffiltext{7}{Departamento de Ciencias Fisicas, Universidad Andres Bello, Campus La Casona, Fernández Concha 700, Santiago, Chile}
\altaffiltext{8}{Millennium Institute of Astrophysics, Av. Vicuña Mackenna 4860, 782-0436 Macul, Santiago, Chile}
\altaffiltext{9}{Vatican Observatory, Vatican City State, I-00120, Italy}
\altaffiltext{10}{Harvard-Smithsonian Center for Astrophysics, 60 Garden Street, Cambridge, MA 01238, USA}
\altaffiltext{11}{Physics Department, University of Michigan, Ann Arbor, MI 48109, U.S.A. ; Astronomy Department, University of Michigan, Ann Arbor, MI 48109, USA}
\altaffiltext{12}{Hubble Fellow}
\altaffiltext{13}{NASA Sagan Fellow}


\newcommand{\about}{$\sim$~}
\newcommand{\mj}{M$_{J}$}
\newcommand{\degrees}{$^{\circ}$}
\newcommand{\arcseconds}{$^{\prime \prime}$}
\newcommand{\asec}{$\arcsec$}
\newcommand{\fasec}{$\farcs$}
\newcommand{\lprime}{$L^{\prime}$}
\newcommand{\ks}{$Ks$~}
\newcommand{\mjyasec}{mJy/arcsecond$^{2}$}
\newcommand{\microns}{$\mu$m}
\newcommand{\msun}{M$_{\odot}$}
\newcommand{\ms}{ms$^{-1}$}
\newcommand{\msini}{$m \sin{i}$} 

\shortauthors{Rodigas et al.}

\begin{abstract}
We present high-contrast Magellan adaptive optics (MagAO) images of HD 7449, a Sun-like star with one planet and a long-term radial velocity (RV) trend. We unambiguously detect the source of the long-term trend from 0.6-2.15 \microns ~at a separation of \about 0\fasec 54. We use the object's colors and spectral energy distribution to show that it is most likely an M4-M5 dwarf (mass \about 0.1-0.2 \msun) at the same distance as the primary and is therefore likely bound. We also present new RVs measured with the Magellan/MIKE and PFS spectrometers and compile these with archival data from CORALIE and HARPS. We use a new Markov chain Monte Carlo procedure to constrain both the mass ($> 0.17$ \msun ~at 99$\%$ confidence) and semimajor axis (\about 18 AU) of the M dwarf companion (HD 7449B). We also refine the parameters of the known massive planet (HD 7449Ab), finding that its minimum mass is $1.09^{+0.52}_{-0.19}$ \mj, its semimajor axis is $2.33^{+0.01}_{-0.02}$ AU, and its eccentricity is $0.8^{+0.08}_{-0.06}$. We use N-body simulations to constrain the eccentricity of HD 7449B to $\lesssim$ 0.5. The M dwarf may be inducing Kozai oscillations on the planet, explaining its high eccentricity. If this is the case and its orbit was initially circular, the mass of the planet would need to be $\lesssim$ 1.5 \mj. This demonstrates that strong constraints on known planets can be made using direct observations of otherwise undetectable long-period companions. 
\end{abstract}
\keywords{instrumentation: adaptive optics --- techniques: high angular resolution --- techniques: radial velocity --- stars: individual (HD 7449) --- binaries --- planetary systems}

\section{Introduction}
Direct imaging and radial velocity (RV) are complementary planet detection techniques. RV is typically sensitive to gas giant planets orbiting within \about 5 AU of old, Sun-like, chromospherically quiet stars. Direct imaging can detect super-Jovian planets orbiting beyond \about 10 AU of young, massive stars. Stars with systems that bridge the desired characteristics of the two methods are thus ideal targets for both RV and imaging. 

The most obvious candidates are stars that show long-term RV trends, which indicate the presence of one or more massive companions on long-period orbits. Because imaging contrast improves far from the star's point spread function (PSF), such objects are ideal targets for imaging. The combined power of RV and direct imaging has been realized on several systems to date. A few M dwarfs have been imaged within 25 AU of stars that also host eccentric planets \citep{gammacep,chauvinrvimaging,gliese86,howardrvimaging}. \cite{rvluckyimaging} directly imaged an M dwarf companion to a star that showed a long-term RV signal and used the derived photometric mass to constrain the system inclination. The TRENDS survey \citep{crepptrends1,crepptrends3,crepptrends2,crepptrends4,crepptrends5} is specifically dedicated to targeting stars that have long-period RV trends. Several stellar and substellar companions have been discovered and characterized, helping to constrain the atmospheres of cool objects. This is especially relevant given the growing number of cool substellar and planetary mass objects being discovered by direct imaging. Even null-detections are useful, as \cite{eind} and \cite{14her} used 4 \microns ~thermal imaging to set strong constraints on the types of substellar companions that could orbit two nearby stars.

We are conducting an adaptive optics (AO) direct imaging survey of nearby southern-hemisphere stars that have long-term RV trends. The stars are selected from the combined RV planet surveys using the AAT/UCLES, Magellan/MIKE \citep{mike}, and Magellan/PFS \citep{pfs} instruments. The imaging is performed using the Magellan adaptive optics system (MagAO, \citealt{magao}), which offers simultaneous high Strehl ratio imaging in the visible (with VisAO, \citealt{visao}) and the infrared (with Clio-2, \citealt{suresh}). The ability to image in the visible is a key advantage compared to other AO-enabled telescopes because an imaged object's spectral energy distribution (SED) can then be constructed in a single night.

In this first paper, we report our observations of the Sun-like star HD 7449 located 38.9$^{+0.74}_{-0.71}$ pc away \citep{updatedhip}. HD 7449 is thought to be a sub-solar metallicity ($[F_{e}/H]$ = -0.11$\pm 0.01$,  \citealt{hd7449dumusque}, in agreement with \citealt{metallicities1} and \citealt{sweetcat}) F8V star. Its age is estimated as 2.10 $\pm$ 0.24 Gyr old \citep{hd7449dumusque} based on the age-activity relations from \cite{mamajek}. \cite{hd7449dumusque} used HARPS and CORALIE RV data to suggest that HD 7449 has a planet with mass $>$ 1.1 \mj ~at 2.3 AU and a long-term trend, which they concluded was most likely arising from a planet with mass $>$ 2 \mj ~at 5 AU. The preferred orbits of these planets were very eccentric. \cite{wit13} searched for solutions containing two planets on near-circular orbits because such systems can often be mistaken for systems with only a single eccentric planet \citep{me,guillemecc}. They preferred solutions of a planet with mass $>$ 1.2 \mj ~at 2.83 AU and a second planet with mass $>$ 0.4 \mj ~at 1.44 AU, both with near-circular orbits. 

Speckle interferometry searches for substellar companions close to the star have to date resulted in null-detections \citep{hd7449speckle}. Using MagAO's simultaneous visible and infrared imaging capabilities coupled with high Strehl ratio AO, we have detected a faint object at a projected separation of \about 0\fasec 54 around HD 7449. In Section 2 we describe our observations, which include both imaging at seven wavelengths from 0.63-2.15 \microns ~and new Doppler spectroscopy, and we describe our data reduction. In Section 3 we present photometry and astrometry for the object and show that it is an M dwarf at the same distance as the primary and thus is likely the source of the long-period trend; we also constrain its mass and period from RV analysis and provide updated parameters on the known inner planet HD 7449Ab; and we use numerical N-body simulations to further constrain the architecture of the system. In Section 4 we discuss the implications of our results, compare HD 7449 to other similar systems, and conclude.

\section{Observations and Data Reduction}
\subsection{MagAO Imaging}
\begin{figure*}[t]
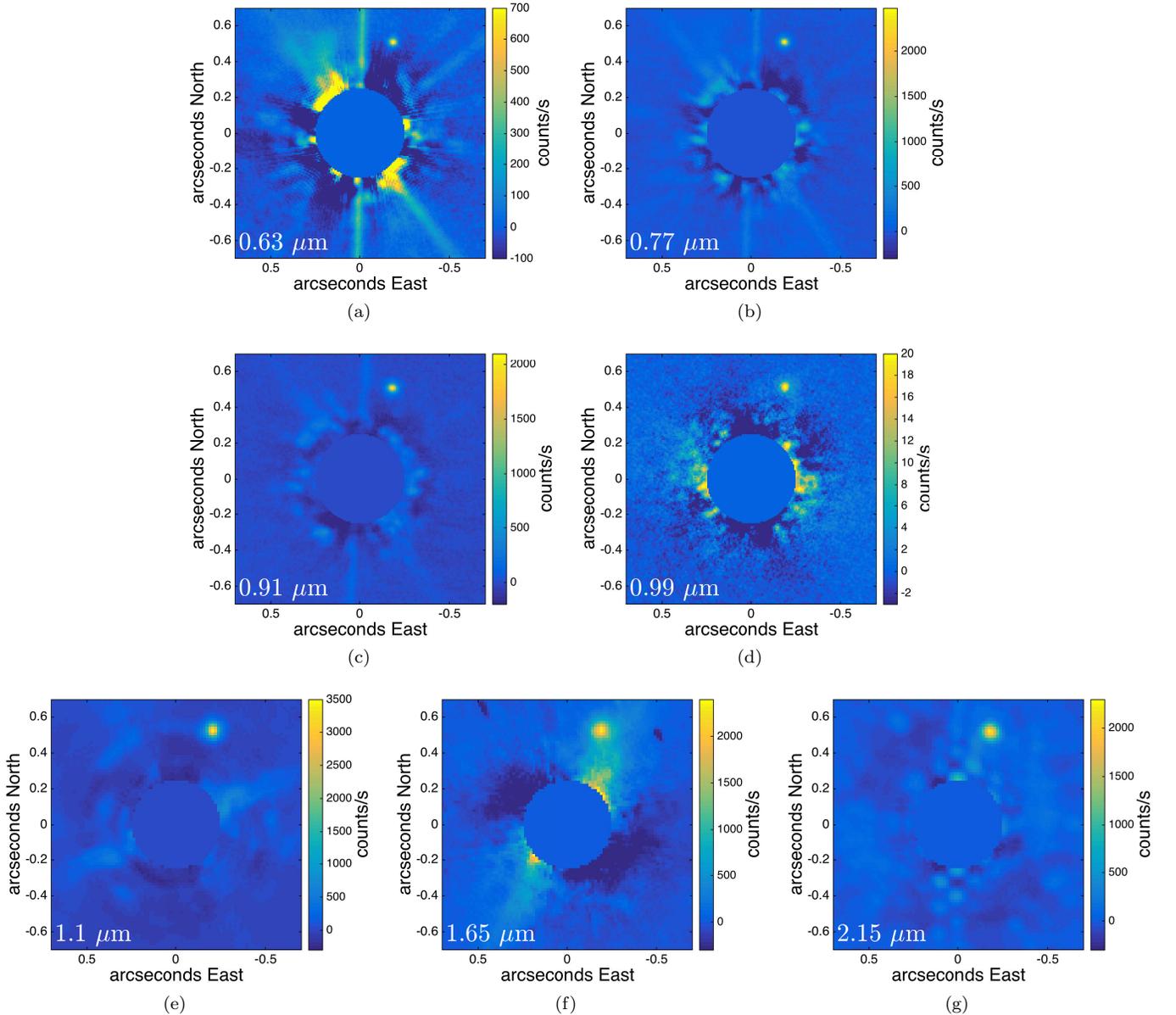

\centering
\subfloat[]{\label{fig:r}\includegraphics[width=0.34\textwidth]{rp.eps}} 
\subfloat[]{\label{fig:i}\includegraphics[width=0.34\textwidth]{ip.eps}} \\
\subfloat[]{\label{fig:z}\includegraphics[width=0.34\textwidth]{zp.eps}} 
\subfloat[]{\label{fig:y}\includegraphics[width=0.34\textwidth]{Ys.eps}} \\
\subfloat[]{\label{fig:J}\includegraphics[width=0.34\textwidth]{J.eps}} 
\subfloat[]{\label{fig:H}\includegraphics[width=0.34\textwidth]{H.eps}} 
\subfloat[]{\label{fig:K}\includegraphics[width=0.34\textwidth]{Ks.eps}} \\
\caption{Final reduced images of HD 7449 and its outer companion at seven photometric bands with central wavelengths noted on the panels: $r'$ (a), $i'$ (b), $z'$ (c), $Ys$ (d), $J$ (e), $H$ (f), and $Ks$ (g). North is up and East is to the left, and a 0\fasec 25 radius digital mask around the star has been added for display purposes. Radial profiles have been subtracted from each image to remove the stellar halos, since no PSF subtraction was performed. The companion is clearly visible at a separation of \about 0\fasec 54 and position angle ($P.A.$) of \about 340\degrees.}
\label{fig:images}
\end{figure*}
We observed HD 7449 using the Magellan Clay Telescope at the Las Campanas Observatory in Chile on the nights of UT November 5, 2014 and November 22, 2014. We used MagAO paired with VisAO and Clio-2, for which we used the narrow camera (plate scale = 0\fasec 01585; \citealt{ktbetapic}). On the first night, the observing conditions were fair, with seeing varying around 1\asec, therefore only 200 modes of AO correction were employed. We observed the star with VisAO at $Ys$ (0.99 \microns) and with Clio-2 at $H$ (1.65 \microns) and $Ks$ (2.15 \microns). Unsaturated photometric images were also acquired in each filter. On the second night, the seeing was much better, with stable seeing under 1\asec, therefore the maximum 300 modes of AO correction were employed. We observed the star with VisAO at $r'$ (0.63 \microns), $i'$ (0.77 \microns), $z'$ (0.91 \microns), and with Clio-2 at $J$ (1.1 \microns). Unsaturated photometric images were acquired in each filter. All observations were acquired with the instrument rotator off to enable angular differential imaging (ADI, \citealt{adi}).

A bright object was identifiable in the raw images at each wavelength, separated by \about 0\fasec 5 from the star. Therefore ADI PSF subtraction was not needed to enhance contrast, and little integration was required in each filter. We obtained total integrations of 2.3 minutes at $r'$, 1.2 minutes at $i'$, 1.17 minutes at $z'$, 1.9 minutes at $Ys$, 0.5 minutes at $H$, 18.67 minutes at $J$, and 4.33 minutes at $Ks$. 

All data reduction was performed with custom scripts in Matlab. The Clio-2 images were divided by the number of coadds, corrected for nonlinearity \citep{ktbetapic}, divided by the integration times, sky-subtracted, and then registered and cropped. The VisAO images were dark-subtracted, divided by the integration times, and then registered and cropped. All images were rotated to North-up, East-left and then median-combined into final images at each wavelength. Finally, 2D radial profiles were subtracted from each image to remove the majority of the stellar flux (see Fig. \ref{fig:images}). The object at \about 0\fasec 54 is unambiguously detected in each filter. 

\subsection{Doppler Spectroscopy}
RV data on HD 7449 were first acquired as part of the Magellan Planet Search Program, which originally made use of the MIKE echelle spectrometer \citep{mike} on the Magellan Clay telescope until September 2009. 
\begin{table}[h]
\centering
\caption{RVs for HD 7449}
\label{tab:RVs}
\begin{tabular}{c c c c}
\hline
\hline
Julian Date & RV (m/s) & $\sigma_{RV}$ (m/s) & Instrument \\
\hline
2451459.55882	&	78.57	&	10.00	&	1 \\
2451480.14706	&	86.90	&	10.00	&	1 \\
2451490.44118	&	76.19	&	10.00	&	1 \\
2451541.91176	&	76.90	&	10.00	&	1 \\
2451747.79412	&	52.62	&	10.00	&	1 \\
... \\
\hline
\end{tabular}  
\end{table} 
\begin{table}[h!]
\vspace{-0.68cm}
\raggedright \textbf{Notes.} Table \ref{tab:RVs} is published in its entirety in the electronic edition of XXX. A portion is shown here for guidance regarding its form and content. Instrument 1 corresponds to CORALIE \citep{hd7449dumusque}, 2 corresponds to HARPS, 3 corresponds to Magellan/MIKE, and 4 corresponds to Magellan/PFS. \\ 
\end{table}
The reported precision achieved by MIKE was 5 \ms ~on solar type stars \citep{mikeprecision}. Observations with MIKE were made using a 0\fasec 35 slit, which results in a spectral resolution of R $\sim$ 70,000 in the blue and $\sim$ 50,000 in the red. The wavelength coverage ranges from 3900 to 6200 $\AA$, capturing the iodine region (5000-6300 $\AA$), and is divided into two CCDs covering the red and blue wavelength regions.

\begin{figure}[t!]
\centering
\subfloat[]{\label{fig:RVs}\includegraphics[width=0.47\textwidth]{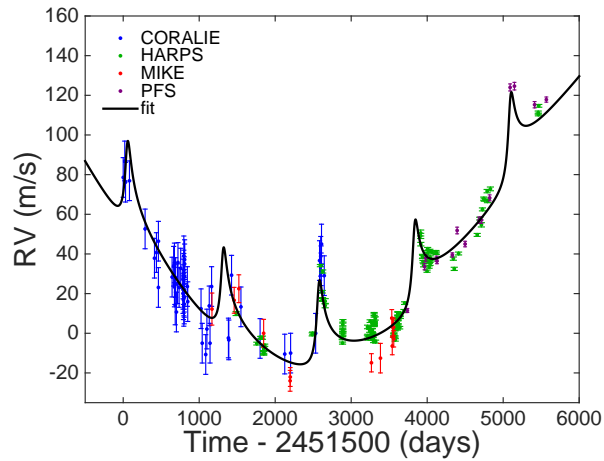}} \\
\subfloat[]{\label{fig:bphased}\includegraphics[width=0.47\textwidth]{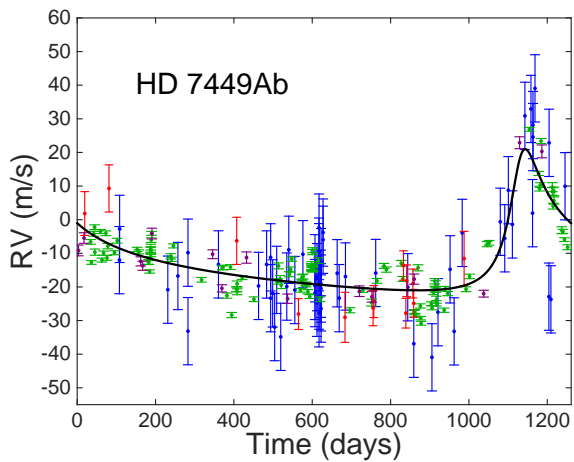}} \\
\subfloat[]{\label{fig:Bphased}\includegraphics[width=0.47\textwidth]{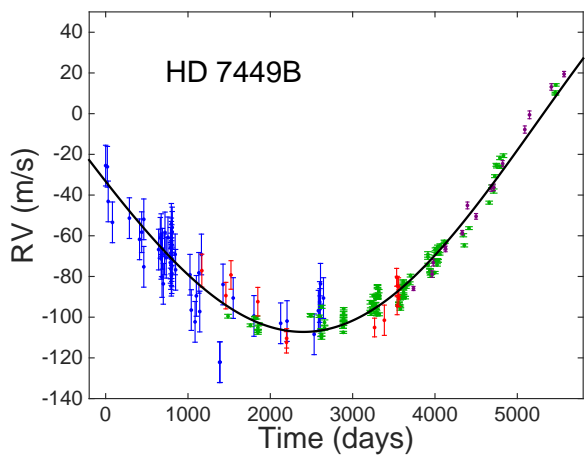}} \\
\caption{RVs for HD 7449. Blue, green, red, and purple points correspond to CORALIE \citep{hd7449dumusque}, HARPS, and Magellan/MIKE and PFS, respectively. (a) The RV data and the combined best-fit (solid black line). (b-c) The phase-folded RV data and fits to the two strongest signals, the massive planet on a very eccentric orbit (HD 7449Ab) and the long-period companion (HD 7449B), with the other signals removed in each case. }
\end{figure}

HD 7449 was subsequently observed using the Carnegie Planet Finder Spectrograph (PFS, \citealt{pfs}), a temperature-controlled high resolution spectrograph, which now carries out all observations for the Magellan Planet Search Program. PFS covers 3880 to 6680 $\AA$ and the 0\fasec 5 slit is used, which results in a spectral resolution of $\sim$ 80,000 in the iodine region. Continuous monitoring of stable stars reveals that the Magellan/PFS system achieves an average measurement precision of 1.5 \ms ~\citep{pfsprecision}.

The RVs for both instruments were obtained using the iodine technique \citep{pauliodine}. Briefly, an iodine absorption cell provides the wavelength scale and instrumental PSF for each stellar observation, which are computed in 2 $\AA$ chunks. A forward modeling procedure of each observation is carried out for each chunk, thus providing an individual measurement of the wavelength, PSF, and Doppler shift. The final measured RV is the weighted average of all the chunks for a given observation. Internal uncertainties are computed as the standard deviation of the velocities derived from each chunk. The new RVs for HD 7449 obtained from MIKE and PFS are listed in Table \ref{tab:RVs}.

We also included in our analysis RVs measured with HARPS and CORALIE. These RVs were originally reported in \cite{hd7449dumusque}. However, HD 7449 has been observed by HARPS since that publication, so we downloaded all available HARPS data on HD 7449 from the ESO archive. Starting from the ESO extracted and calibrated spectra, we obtained new Doppler measurements using the HARPS-TERRA software \citep{terra}. The CORALIE data were not explicitly reported by \cite{hd7449dumusque}, nor are they available in any archive, so we used \textit{DataThief} (\url{http://datathief.org}) to retrieve the RVs. To account for possible errors in the extraction, we assumed 10 \ms ~errors for the CORALIE data in our subsequent RV analysis. The entire RV data set is shown in Fig. \ref{fig:RVs}, revealing the clear long-term, parabolic trend, and the individual RVs are reported in Table \ref{tab:RVs}.


\section{Results}
\subsection{Outer Companion Photometry and Astrometry}
Photometry was measured as follows. First, a circular aperture of radius = 1 full-width half-maximum (FWHM), corresponding to the size of a diffraction-limited PSF at each wavelength, was placed at the detected object's photocenter in each image. The same aperture was placed at the stellar photocenter in each unsaturated image, and then the fluxes within all the apertures were summed. Uncertainties were calculated as the standard deviations of the fluxes in apertures placed around the star at the same radius. 
\begin{figure}[t]
\centering
\includegraphics[width=0.49\textwidth]{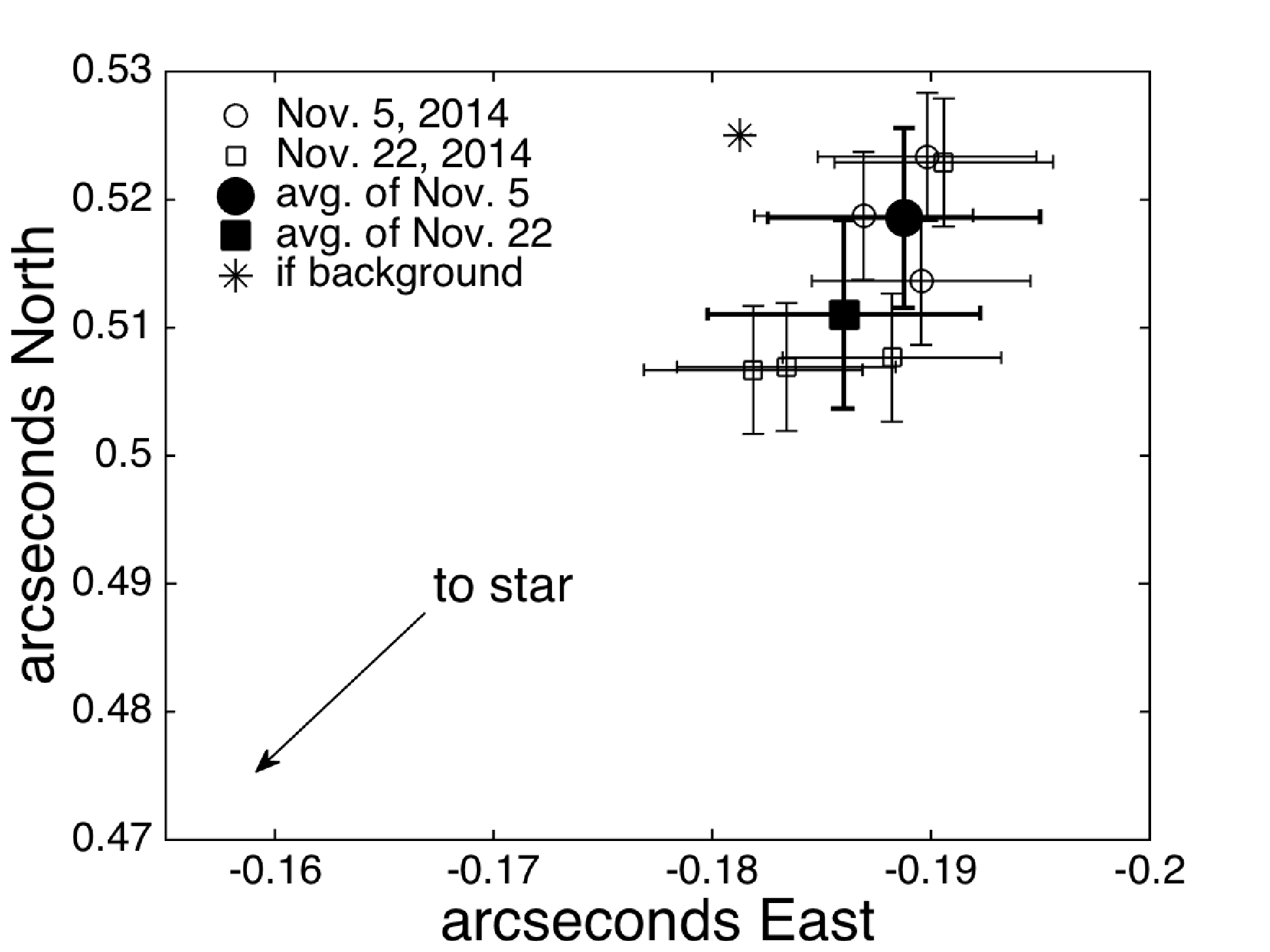}
\caption{Astrometry of HD 7449B from our two epochs of MagAO imaging. The circles correspond to the detections on Nov. 5, 2014 and the squares correspond to Nov. 22, 2014. The asterisk denotes where the companion would have been located on Nov. 22, 2014 if it were a background object, based on the star's proper motion \citep{updatedhip}. The object's motion over 17 days is inconsistent with a background object at the 2$\sigma$ confidence level.}
\label{fig:astrometry}
\end{figure}

\begin{table}[t]
\centering
\caption{HD 7449B Photometry and Astrometry}
\label{tab:phot}
\begin{tabular}{c c}
\hline
\hline
Parameter & Value \\
\hline
$\Delta r'$ (0.63 \microns) & 8.82$^{+0.13}_{-0.11}$ \\
$\Delta i'$ (0.77 \microns) & 7.32$^{+0.13}_{-0.11}$ \\
$\Delta z'$ (0.91 \microns) & 6.53$^{+0.15}_{-0.13}$ \\
$\Delta Ys$ (0.99 \microns) & 5.87$^{+0.29}_{-0.23}$ \\
$\Delta J_{MKO}$ (1.1 \microns) & 5.81$^{+0.11}_{-0.10}$ \\
$\Delta H_{MKO}$ (1.65 \microns) & 5.11$^{+0.11}_{-0.10}$ \\
$\Delta Ks_{Barr}$ (2.15 \microns) & 4.85$^{+0.03}_{-0.03}$ \\
$M_{r'}$ & 13.39$\pm$0.17  \\ 
$M_{i'}$ & 11.51$\pm$0.17    \\
$M_{z'}$ & 10.75$\pm$0.20 \\
$M_J$ & 9.26$\pm$0.16  \\
$M_H$ & 8.33$\pm$0.16 \\
$M_{K_s}$ & 7.97$\pm$0.09 \\
\hline 
$\Delta RA_{t_1}$ (\asec) & -0.19$\pm 0.003$ \\
$\Delta Dec_{t_1}$ (\asec) & 0.52$\pm 0.003$ \\
$\Delta RA_{t_2}$ (\asec) & -0.19$\pm 0.006$ \\
$\Delta Dec_{t_2}$ (\asec) & 0.51$\pm 0.005$ \\
$\rho_{t_1}$ (\asec) & 0.55$\pm 0.007$ \\
$P.A._{t_1}$ (\degrees) & 339.99$\pm 1.84$ \\
$\rho_{t_2}$ (\asec) & 0.54$\pm 0.007$ \\
$P.A._{t_2}$ (\degrees) & 339.99$\pm 1.88$ \\
\hline
\end{tabular}  
\end{table} 
\begin{table}
\vspace{-0.35cm}
\raggedright \textbf{Notes.} $t_1$ = UT Nov. 5, 2014; $t_2$ = UT Nov. 22, 2014. $M_{Ys}$ is not reported (or used in any photometric analysis) because the primary star has no reported measurements near 1 \microns. \\ 
\end{table}

Astrometry was measured by calculating the photocenters in the same apertures, and astrometric uncertainties were assumed to be 5 mas at each wavelength based on previous imaging with MagAO (e.g., \citealt{me4796}). Table \ref{tab:phot} lists the object's photometry and astrometry. The object has a separation of \about 0\fasec 54 and $P.A.$ \about 340\degrees. Because the star has high proper motion \citep{updatedhip}, the two epochs of direct detections separated by only 17 days is enough to show that the object is inconsistent with being background at 2$\sigma$ confidence. In Section \ref{sec:mass}, we will show that the object's SED confirms that it is unlikely to be background. 

Henceforth, we will refer to the outer object as HD 7449B. Note that \cite{hd7449binary} suggest that HD 7449 has a common proper motion companion at $>$ 2000 AU. The candidate companion was identified using the PPMXL proper motion catalog \citep{ppmxl}. Examining the relevant images from PPMXL reveals that the object is actually one of the diffraction spikes and is therefore not a real astrophysical source. Therefore HD 7449 does not have any stellar companions at $> 2000$ AU.


\subsection{Outer Companion Mass from Photometry}
\label{sec:mass}
Because we have detections of HD 7449B in both the visible and the near-infrared (NIR), we can use its colors and absolute magnitudes to constrain its spectral type, effective temperature ($T_{eff}$), and mass. To accomplish this, we compared its photometry to both known objects and to the low-mass stellar models of \cite{krauscoolstars} and Baraffe (\citealt{baraffe98,baraffe02,baraffe15}). 

\begin{figure}[t]
\centering
\includegraphics[width=0.49\textwidth]{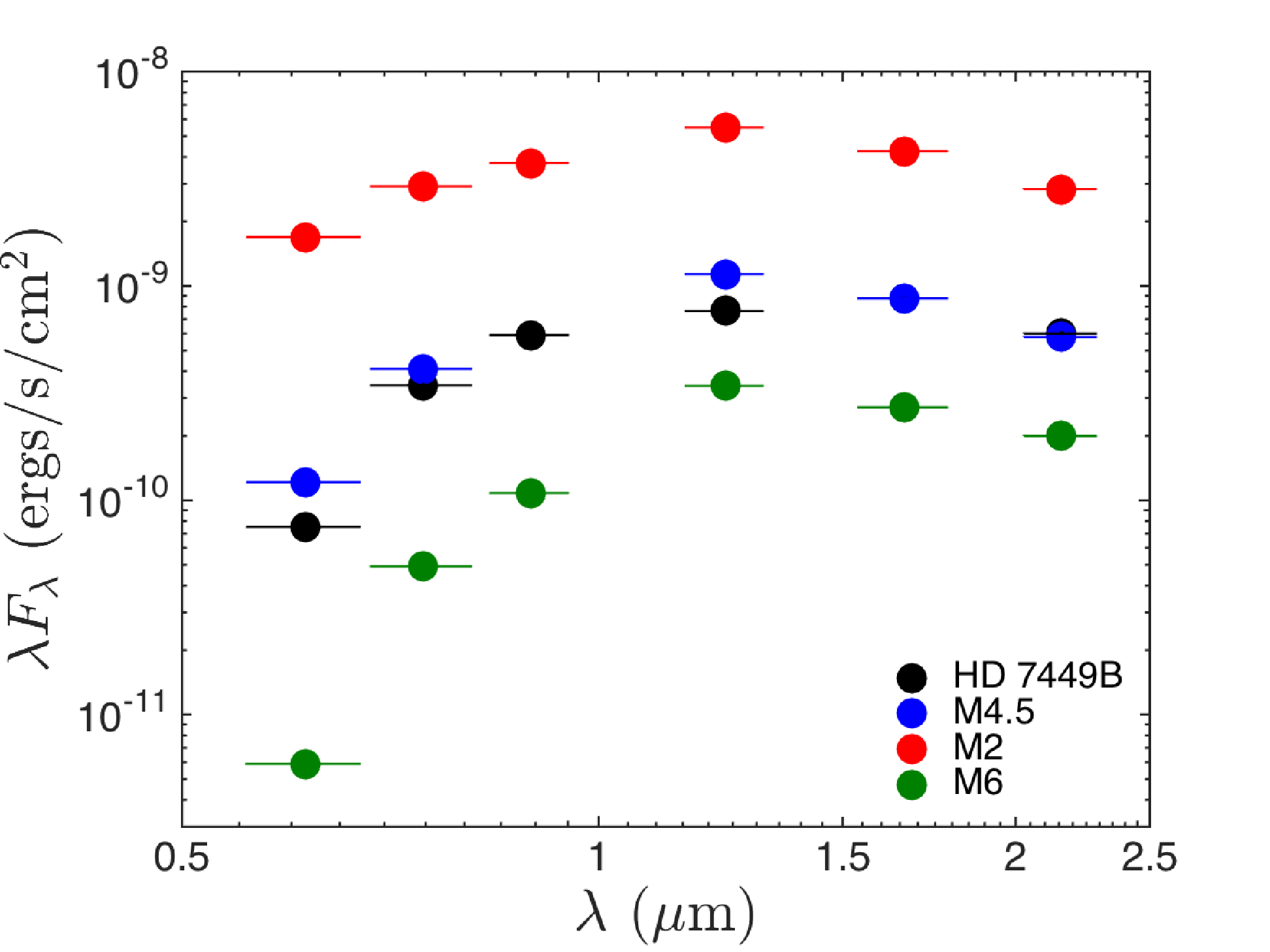}
\caption{SED of HD 7449B, along with other M dwarfs. Error bars are smaller than the marker sizes. The companion's SED point at $Ys$ is not shown because HD 7449A has no measured flux at this wavelength. The companion's SED point at $H$ lies behind the point corresponding to the M4.5, which itself has no $z'$ flux measurement. HD 7449B is most similar to the M4.5 source. This also confirms that it is likely to be at the distance to the primary (38.9 pc).}
\label{fig:sed}
\end{figure}

To create a comparative SED for HD 7449B, we began with the MagAO photometry for the primary and the outer companion. We used catalog 2MASS and SDSS photometry for HD 7449A and then used the color transformation relations in \cite{colortransforms} to put the NIR photometry on the MKO system, which is comparable to the MagAO filters. Photometry for HD 7449B was then obtained by computing the magnitude differences relative to the primary. We used the Hipparcos parallax of 25.69$\pm$0.48 mas \citep{updatedhip} to compute the absolute magnitudes and then converted each to $\lambda F_{\lambda}$ (e.g., see \cite{jackie13}). Fig. \ref{fig:sed} shows the resulting SED for HD7449B as well as the similarly-computed SEDs of comparative M dwarfs from the 8-parsec sample \citep{nearbymdwarfs}. The best matching SED corresponds to an M4.5, which also confirms that HD 7449B is at the distance to the primary (38.9 pc).

\begin{figure*}[t]
\centering
\subfloat[]{\label{fig:hkh}\includegraphics[width=0.49\textwidth]{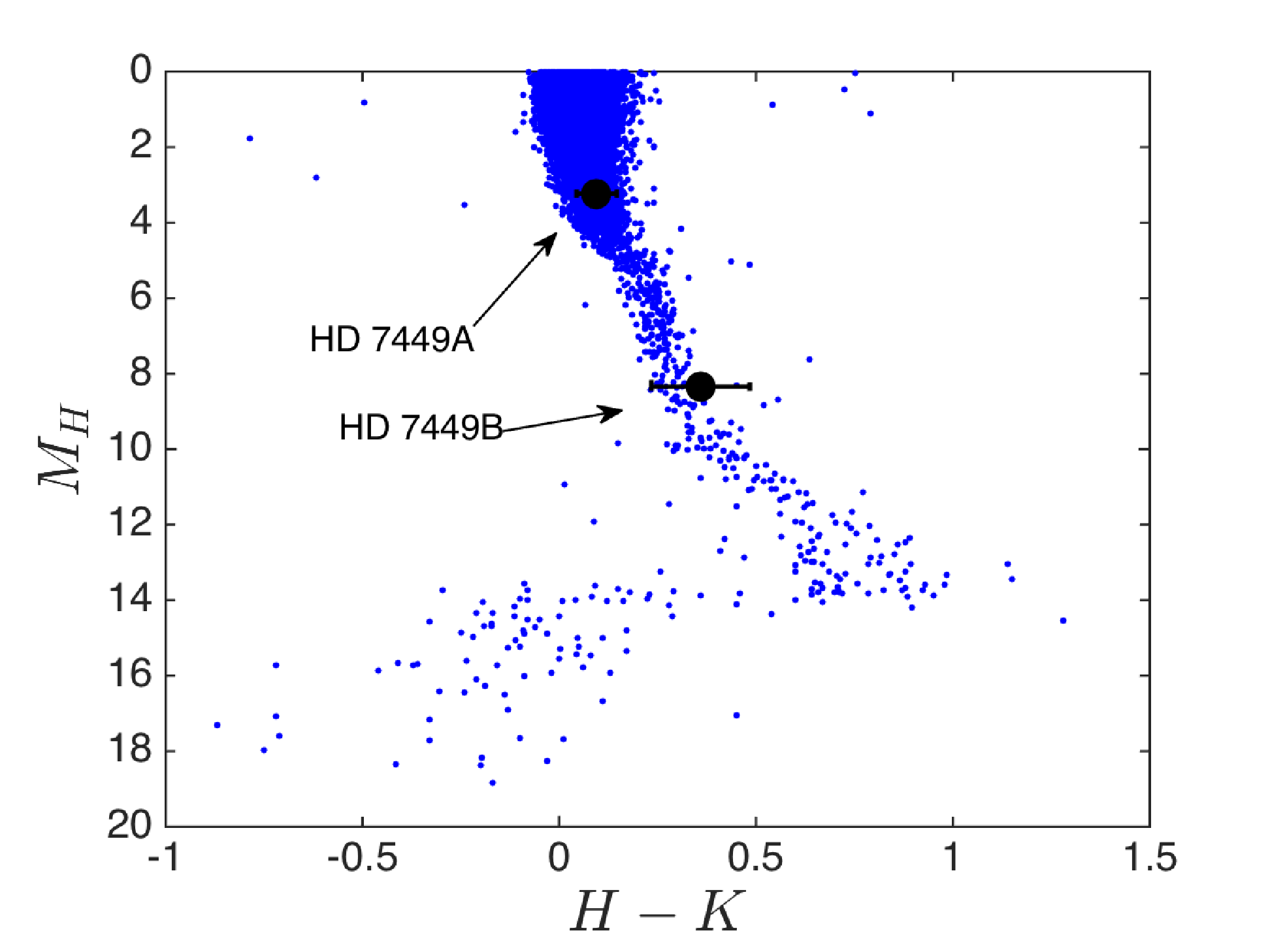}} 
\subfloat[]{\label{fig:hkk}\includegraphics[width=0.49\textwidth]{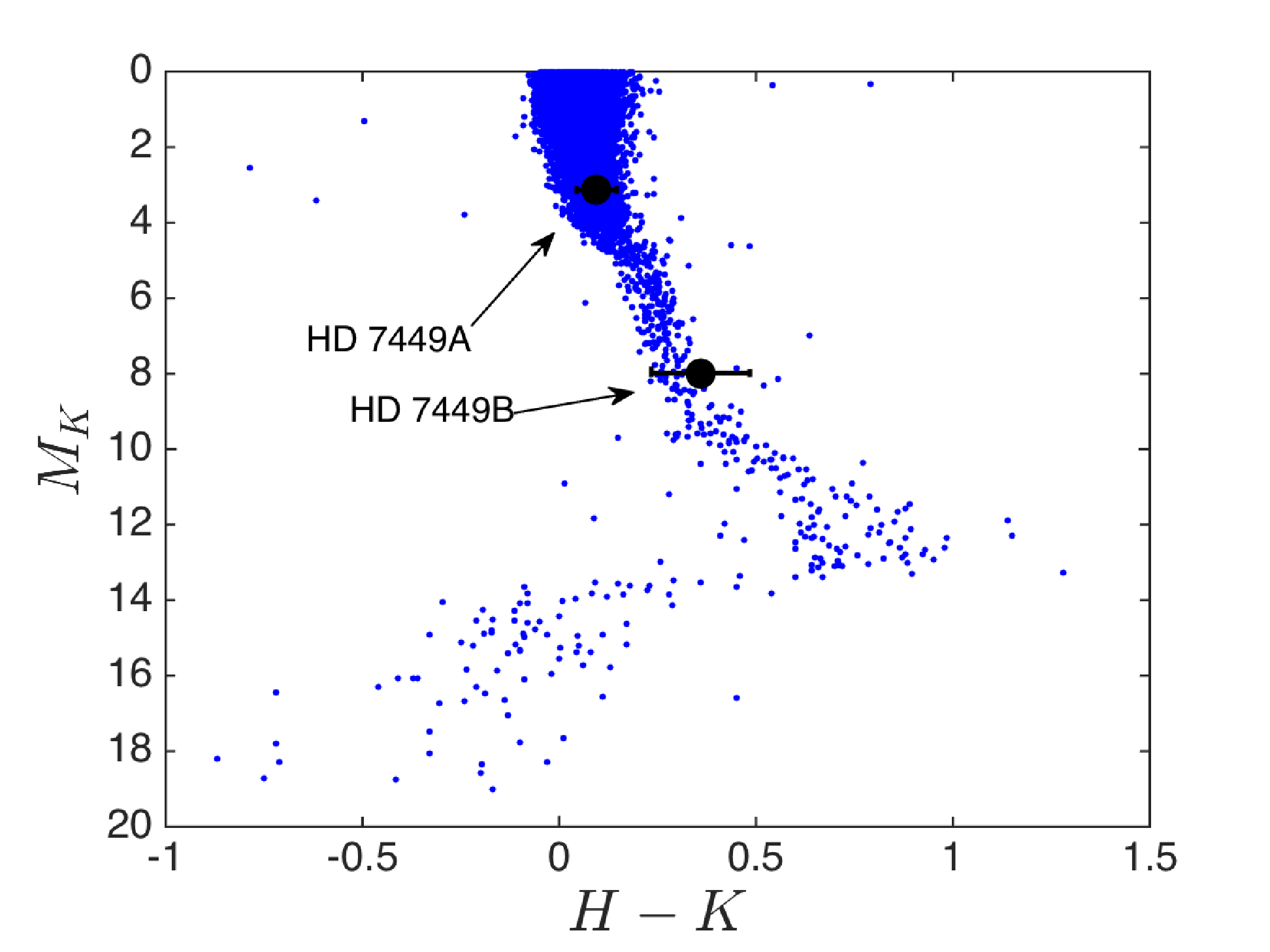}} \\
\caption{NIR CMDs for HD 7449A and B (black points). Blue points correspond to cool stars and brown dwarfs from the Hipparcos, NSTARS, \cite{trent12}, and \cite{jackie12} samples. These CMDs show that the A and B components fall on the main sequence together, which means they are likely to be co-eval. Therefore HD 7449B is unlikely to be a background object.}
\label{fig:cmd}
\end{figure*}

To demonstrate that HD 7449A and B fall along the main sequence together (hence verifying that they are likely co-eval), we constructed color magnitude diagrams (CMDs) at several wavelengths from the visible to the NIR. We used the low-mass star Hipparcos sample, the NSTARS parallax sample, and the brown dwarf parallax sample from \cite{trent12} and \cite{jackie12}. Because these report photometry in the 2MASS system, we converted our MagAO photometry to 2MASS (assuming MKO comparable) using the \cite{colortransforms} relations. All CMDs generally showed that the A and B components fall on the main sequence together, indicating that they are co-eval and that the companion is not a background or foreground object. Fig. \ref{fig:cmd} shows an example CMD.

Using the \cite{krauscoolstars} models, comparing HD 7449B's colors and absolute magnitudes yielded a best-matching spectral type of M5, $T_{eff} = 3010 K$, and $M = 0.15 M_{\odot}$. Using the pre-2015 Baraffe models \citep{baraffe98,baraffe02}, and assuming the stellar age is between 1-3 Gyr, the colors were best matched by a 0.10 $M_{\odot}$, $T_{eff} = 2824 K$, 1 Gyr old star. The absolute magnitudes were best matched by a 0.15 $M_{\odot}$, $T_{eff} = 3161 K$, 1 Gyr old star. Using the 2015 Baraffe models, for stellar ages between 1-3 Gyr,  the colors were best matched by a star with $M$ = 0.20 \msun, $T_{eff} = 3261 K$, and the absolute magnitudes were best matched by a star with $M =$ 0.09 \msun ~and $T_{eff} = 2643 K$. Based on all of the above analysis, we classify HD 7449B (from photometry alone) as an M4.5$\pm 0.5$ with mass = $0.15\pm 0.05 M_{\odot}$.


\subsection{Constraints from RV Fitting}
\label{sec:rv}
RVs have been obtained on HD 7449 for the past \about 15 years by HARPS and CORALIE \citep{hd7449dumusque}, and by Magellan/MIKE and PFS (this work; see Table \ref{tab:RVs}). To explain the periodic RVs (and the clear long-term trend), previous works \citep{hd7449dumusque,wit13} searched for solutions explained by one or more planets. We have the advantage that we know from direct imaging that the system contains an \about M4.5 companion whose current projected separation is $\gtrsim$ 21 AU. Can this companion explain the long-term trend and in doing so help revise the parameters of the inner planet(s)?

To test this, we first analyzed the RVs using log-likelihood periodograms \citep{rvperiodograms,kapteyn} for preliminary period detection and confidence evaluation. Then we used a Bayesian Markov Chain Monte Carlo (MCMC) approach to produce posterior distributions of the allowed parameter values \citep{fordmcmc}. The likelihood function $L$ contains the Keplerian model and a handful of nuisance parameters to account for the arbitrary zero-points of each RV instrument and the different levels of instrumental excess noise (also called jitter, which typically contains the contribution from stellar activity). The likelihood function is given by
\begin{eqnarray}
L &=& \prod_{I}\prod_i^{N_{obs}} l_{i,I}\\
l_{i,I} &=& \frac{1}{\sqrt{2\pi}}\frac{1}{\sqrt{\epsilon_{i}^2+ s_I^2}}
\exp\left[
    -\, \frac{1}{2} \frac{\left(v_{i,I}-v(t, I)\right)^2}{
         \epsilon_{i,I}^2+ s_I^2
	 }
\right]\\
v_{i,I} &=& \gamma_I + \sum_p u(\hat{\kappa}_p; t) \nonumber \\ 
&+& \dot{v}_r(t-t_0)
+ \frac{1}{2}\ddot{v}_r\left(t-t_0\right)^2,\label{eq:rvmodel}
\end{eqnarray}
where $i$ indexes the observations acquired with the $I$th instrument, $\epsilon_{i,I}$ is the nominal uncertainty of each RV measurement, $\gamma_I$ and $s_I$ are the zero-point and extra noise parameters (also called jitter) of each instrument, and the Doppler signal from a companion on the star is encoded in the model $u(\hat{\kappa}_p; t)$, which is a function of time $t$ and the Keplerian parameters $\hat{\kappa_p}$. The Keplerian parameters of the p$^{th}$ companion in the system are: the orbital period $P_p$ (in days), the semi-amplitude $K_p$ (in m/s), the mean anomaly $\mu_{0,p}$ at the reference epoch $t_0$ (in degrees), the eccentricity $e_p$, and the argument of periastron $\omega_p$. The second and third terms in Eq. \ref{eq:rvmodel} account for the possible presence of a long-period candidate whose orbit is only detected as a trend (acceleration, $\dot{v}_r$) plus some curvature (jerk, $\ddot{v}_r$). These two terms are especially important for the analysis that follows.

When performing the Bayesian MCMC analysis, one needs to specify some prior distributions for these parameters. In this paper, we use uniform distributions for the angles $\mu_0$ and $\omega$, as any value would be equally likely a priori. Given that the objects involved are rather massive and the signals large, we also allow for a uniform eccentricity distribution between [0,0.95). For $K_p$, $\gamma_I$, $s_I$, $\dot{v}_r$, and $\ddot{v}_r$, we assume unbound non-normalized uniform priors. While this can cause issues when normalizing the posterior, we are only using the MCMC analysis to sample the \textit{shape} of the posterior, so precise values of the bounds and the normalization factors are unnecessary. Furthermore, because we will later try to constrain the signal with a period much longer than the span of the observations, the possible values of these three quantities will be correlated, so bound priors might eliminate many long period solutions that would otherwise be highly probable. For example, a large $K$ in general requires a large $\gamma$ unless the companion is precisely crossing the plane of the sky at $t_0$ (which is highly unlikely).

Regarding the prior on the period, in this work we assume that the prior is uniform in $1/P$ (equivalent to uniform in frequency). This is motivated by the following. When analyzing time series, the local solutions to periodic signals are approximately equally-spaced in frequency. For example, if one produces a Lomb-Scargle periodogram of a time series and plots period versus power, one will quickly appreciate that the peaks become much broader with increasing period \citep{scargle}. However, when making the same plot in frequency, the peaks appear uniformly distributed over the possible frequencies. As discussed in \cite{tuomi}, in Bayesian statistics the choice of the parameter automatically imposes implicit priors on all other alternative parameterizations. In the case of the period, a uniform prior at very long periods can outweigh the information content on the likelihood, producing a biased result. While this issue is not very severe for periods shorter than the time-span of the observations (typical RV planet search domains), the disrupting effects of the uniform prior become serious if one attempts to constrain very long orbits and becomes strongly dominated by the chosen period cut-off. On the other hand, the frequency parameter does not suffer from such singularities (all very long periods become packed in a single likelihood maxima close to 0) and preserves the role of the likelihood function as the most informative element in the posterior distribution.

Our MCMC algorithm is based on the one described in \cite{fordmcmc}, which uses a Gibbs sampler with independent Gaussian jump functions for each parameter. For each parameter, the proposal function of our Gibbs sampler depends on a scale parameter that needs to be tuned to ensure acceptance rates between 10$\%$ and 30$\%$. This is automatically done by tuning all the scale parameters until they reach the aforementioned acceptance rates (burn-in period). These samples typically amount for $10^6$ iterations and they are not used for the final MCMC analysis. In this paper we only focus on the detection and characterization of the two most significant signals in the RV data (HD 7994Ab and the long-period trend). While there have been other claims of possible candidates in the system \citep{hd7449dumusque,wit13}, we suspect these were artifacts caused by sampling issues with the rather eccentric orbit of HD 7994Ab and the presence of the long-period parabolic trend. 

\subsection{Two Planet MCMC}
\label{sec:firstmcmc}
\begin{figure}[t]
\centering
\subfloat[]{\label{fig:plong}\includegraphics[width=0.49\textwidth]{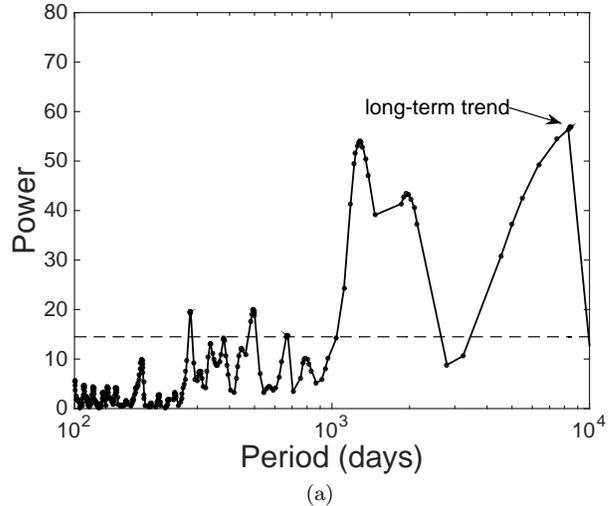}} \\
\subfloat[]{\label{fig:pinner}\includegraphics[width=0.49\textwidth]{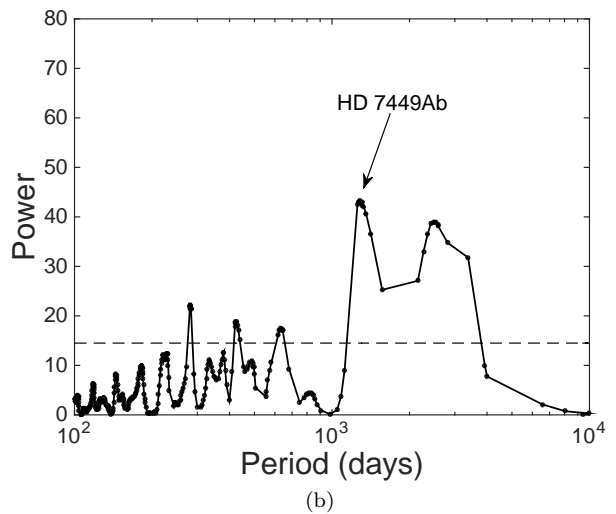}} \\
\caption{Periodograms for the HD 7449 RV data. The dashed line corresponds to the 1 percent false-alarm probability threshold. (a) First periodogram used to identify the strongest signal, which corresponds to the long-period companion. Its period is likely to be $>$ 8000 days. (b) Second periodogram used to identify the next strongest signal at \about 1200 days, corresponding to the previously-identified HD 7449Ab.}
\label{fig:periodograms}
\end{figure}

Our first analysis consisted of a likelihood function with two Keplerian signals: one initialized at P$\sim 1200$ days (which roughly corresponds to the preferred period for the most significant planet in \citealt{hd7449dumusque} and \citealt{wit13}), and the other one at P$\sim$ 8000 days, as suggested by the maximum likelihood periodograms in Fig. \ref{fig:periodograms}. While a maximum likelihood orbit could be obtained with a second planet at $\sim$ 10,000 days (30 years), long MCMC runs indicated that the possible parameters of this object were heavily correlated. As a result, the parameter space was broadly unconstrained, making it difficult for the chains to achieve convergence even after 10$^8$-10$^9$ steps. Such strong degeneracy indicates that only a subset of the 5 Keplerian parameters can be constrained by the current data. As we will see later, the trend in the RV data can be well-described by two terms: an acceleration (linear trend) + jerk (curvature).

\subsection{Two Planet MCMC with Imaging Constraints}
As an attempt at better constraining the orbital elements of the outer companion, in our second analysis we included a Keplerian model for the outer companion's predicted orbital separation in order to use our direct imaging constraints (e.g., see \citealt{binaryorbitmasses}). Unfortunately, the orbital motion of the imaged companion was not very large between the direct imaging runs, and the imaging provides just two observables (projected separation in RA and Dec) while introducing three more free parameters (orbital inclination $i$, longitude of ascending node $\Omega$, and the mass ratio between the companion and the primary star). As a result, these MCMC chains had even more difficulty converging to a meaningful equilibrium distribution (e.g., companion masses up to 100 \msun ~and periods up to millions of years were consistent with the data).

\subsection{One Planet MCMC + Long-period terms}
\label{sec:finalmcmc}
Given that the entire RV data set (including HARPS, CORALIE, MIKE, and PFS RVs) can be well-fit by a simple parabola, for our third analysis we implemented a Doppler model containing a single inner planet plus a linear and quadratic term. In this case, the model contains a single Keplerian initialized at $1200$ days (inner planet), plus the last two terms in Eq. \ref{eq:rvmodel}. Since $\dot{v}_r$ and $\ddot{v}_r$ are linear parameters, the MCMC quickly converged to the best-fit solution, which had an almost identical value of the likelihood function to the full two Keplerian solution attempted in Section \ref{sec:firstmcmc}. This means that the entire RV data set is best (and most simply) described by a single inner planet along with a long-term trend consisting of linear plus quadratic terms. We therefore use this final MCMC's results to constrain the parameters of the companions around HD 7449.

\begin{figure*}[t]
\centering
\subfloat[]{\label{fig:mdist}\includegraphics[width=0.49\textwidth]{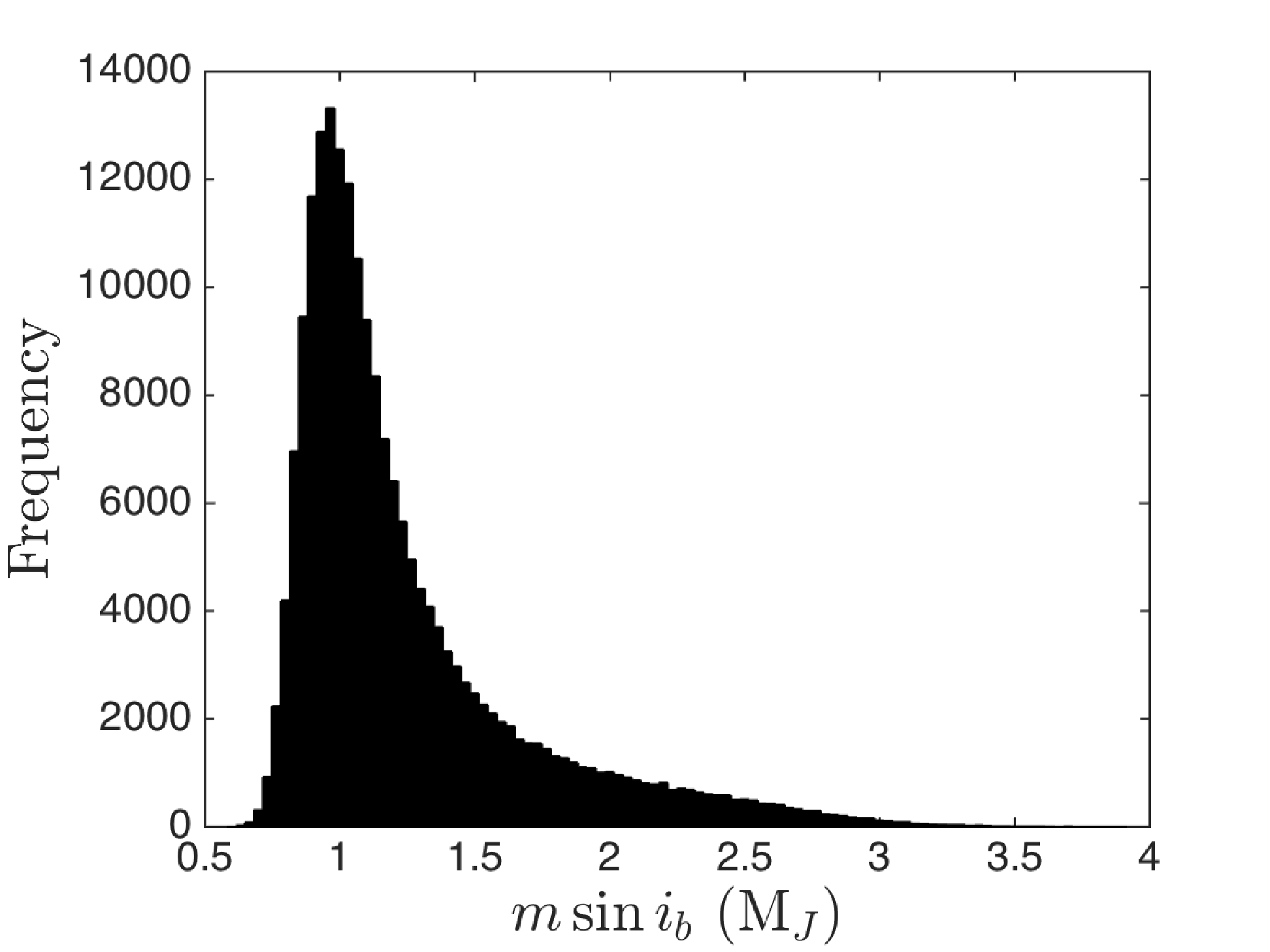}} 
\subfloat[]{\label{fig:Pdist}\includegraphics[width=0.49\textwidth]{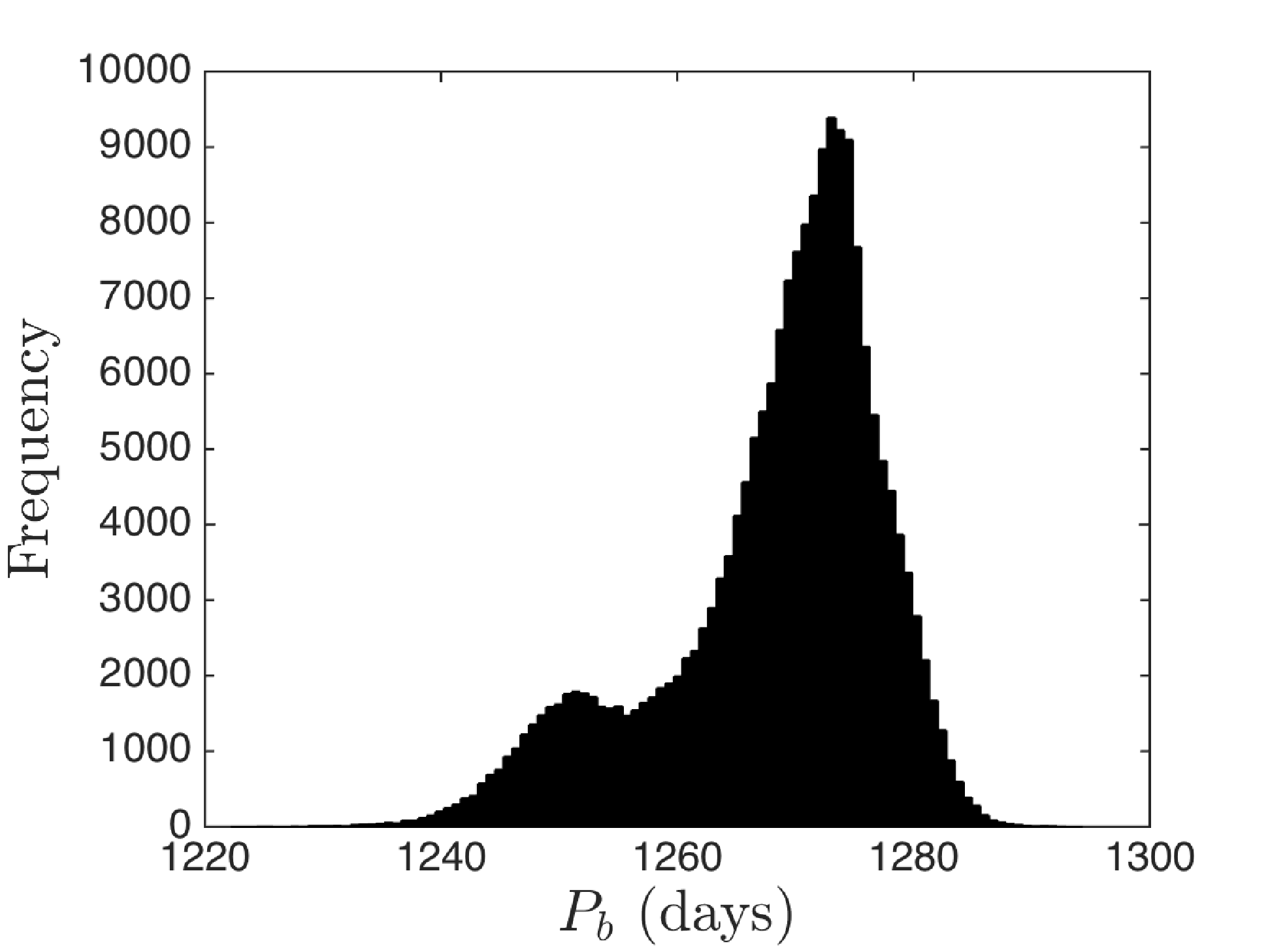}} \\
\subfloat[]{\label{fig:edist}\includegraphics[width=0.49\textwidth]{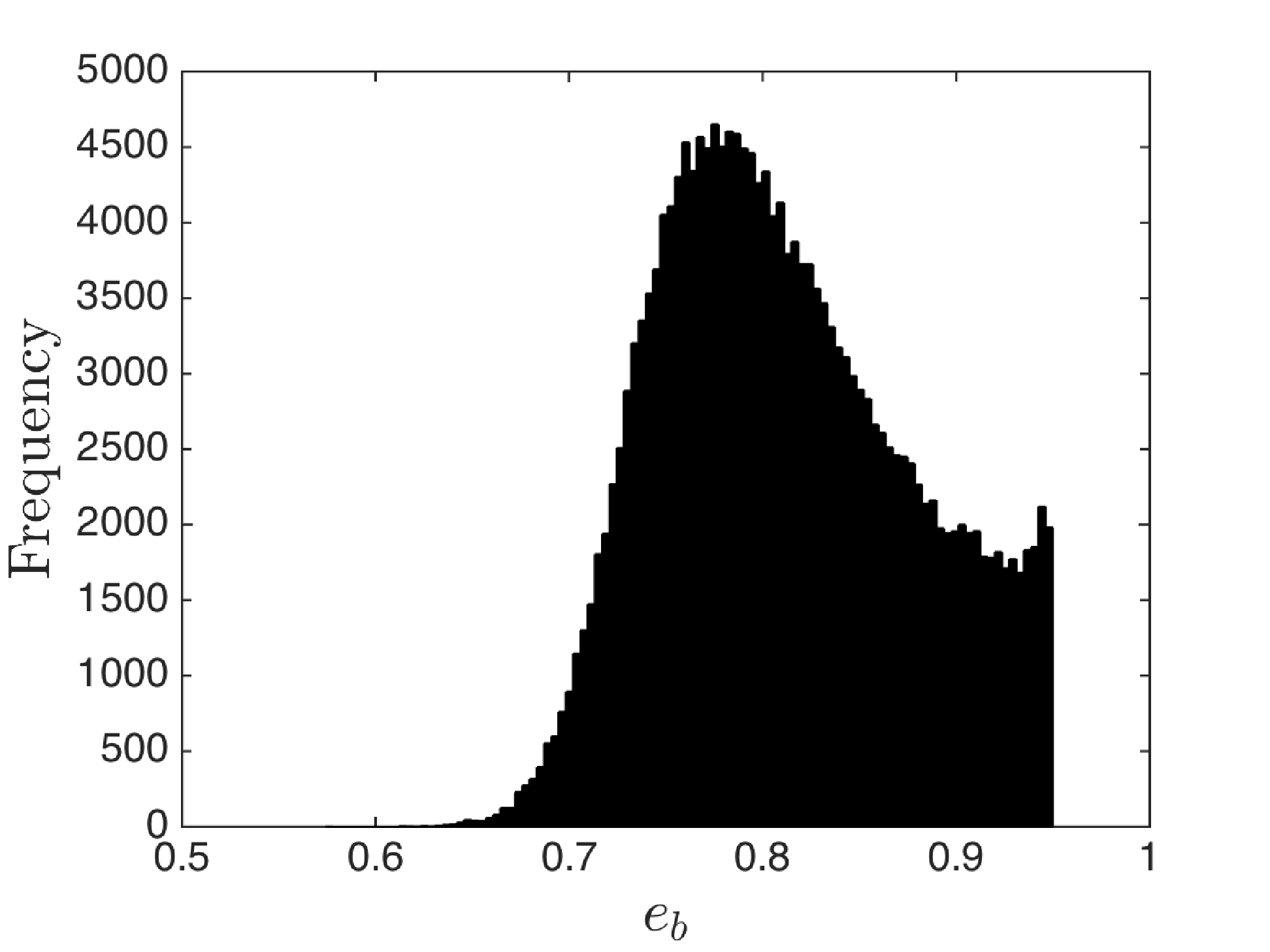}} 
\subfloat[]{\label{fig:wdist}\includegraphics[width=0.49\textwidth]{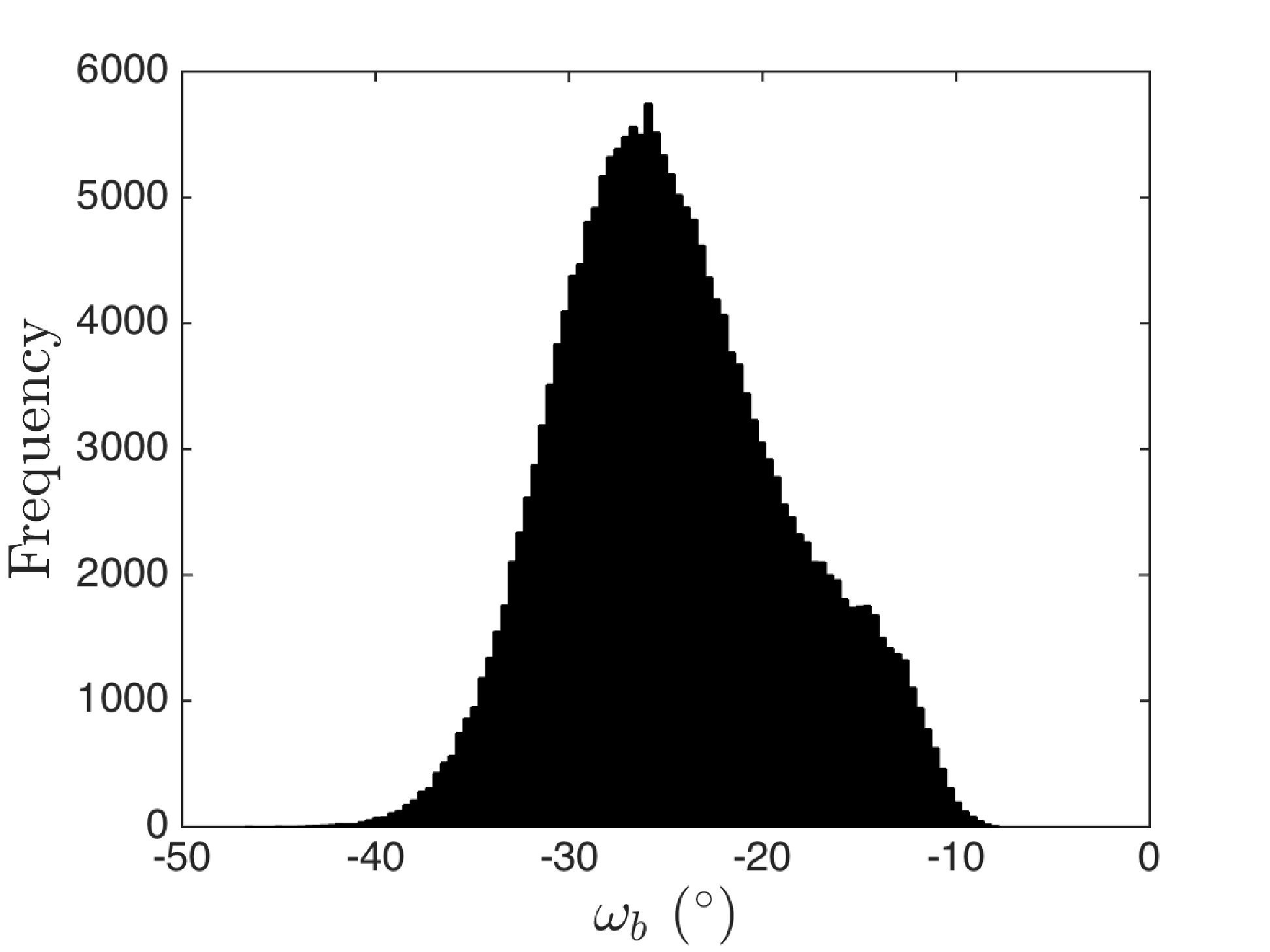}} 
\caption{Marginalized posterior distributions of parameters for HD 7449Ab from our MCMC analysis (Section \ref{sec:finalmcmc}). (a) The planet's minimum mass, $m\sin{i_b}$. (b) The planet's period, $P_b$. (c) The planet's eccentricity, $e_b$. (d) The planet's argument of periastron, $\omega_b$. The planet's properties are tightly constrained: it is likely to be massive and very eccentric, perhaps indicating previous or ongoing dynamical interactions with the outer M dwarf companion (HD 7449B).}
\label{fig:dist1}
\end{figure*}

The posterior distributions of the inner planet's parameters are shown in Fig. \ref{fig:dist1}. Clearly, HD 7449Ab is eccentric (median $e_b = 0.8$), in agreement with \cite{hd7449dumusque}. To constrain the planet's mass ($m$), we used the distributions of $K_b$, $P_b$, and $e_b$, drew random Gaussian-distributed values for the stellar mass $M_{*}$ having mean = 1.05 \msun ~and standard deviation = 0.09 \msun\footnote{\cite{hd7449dumusque} do not report an uncertainty on the stellar mass. Therefore we computed the average of four reported mass values and errors from \cite{sweetcat}, \cite{hd7449mass1}, \cite{hd7449mass2}, and \cite{hd7449mass3}.}, assumed $M_{*} \gg m$, and then solved for $m \sin{i_b}$ using the well-known relation
\begin{equation}
m\sin{i_b} = K_b \sqrt{1-e_b^{2}}  M_{*}^{2/3}  \left(\frac{P_b}{2 \pi G}\right)^{1/3}.
\end{equation}
We find that the median $m\sin{i_b}$ ~= 1.09 \mj, which is in excellent agreement with the preferred mass found by \cite{hd7449dumusque}. HD 7449Ab's parameters and their possible ranges are listed in Table \ref{tab:params}. 

\begin{table*}[t]
\centering
\caption{HD 7449A Companion Parameters}
\label{tab:params}
\begin{tabular}{c|c c c c c c}
\hline
\hline
 & Mass & Period & $a$ (AU) & $e$ & $\omega$ (\degrees) & $i$ (\degrees) \\
\hline
HD 7449Ab & $>1.09^{+0.52}_{-0.19}$ \mj & $1270.5^{+5.92}_{-12.1}$ days & $2.33^{+0.01}_{-0.02}$ & $0.80^{+0.08}_{-0.06}$ & $-25.2^{+6.87}_{-5.22}$ & unconstrained  \\
HD 7449B & $0.23^{+0.22}_{-0.05}$ \msun & $65.7^{+227}_{-56}$ years & $17.9^{+32}_{-12.9}$ & unconstrained  & unconstrained & $59.7^{+20.1}_{-25.8}$ \\
\hline
\end{tabular}  
\end{table*} 
\begin{table*}
\vspace{-0.30cm}
\textbf{Notes.} All uncertainties correspond to symmetric 68$\%$ confidence intervals around the median values. \\ 
\end{table*}

\subsection{Statistical Constraints on Outer Companion}
\label{sec:stats}
Here we develop and apply a new statistical procedure, expanding on the one developed in \cite{rvimagingconstraints}, that uses the slope and quadratic terms discussed in Section \ref{sec:finalmcmc} to tightly constrain the outer companion's properties.

For the case of an imaged companion producing a long-period RV trend, \cite{rvimagingconstraints} formulated a numerical Monte Carlo approach to marginalize over unknown parameters under some uninformative priors. It is based on using the fact that the linear trend observed in a Doppler curve of the primary star can be written in terms of the mass of the long-period companion only ($M_B$), and that the observed separation at a given epoch $t_0$ can be written as a function of $M_B$, the direct imaging observables, and a function that can be easily marginalized over the unknown orbital parameters ($P_B$, $e_B$, $\omega_B$, $\mu_{0,B}$, $i_B$, $\Omega_B$). The method described in \cite{rvimagingconstraints} uses only the measured linear part of the trend and produces a distribution of possible masses. We now develop a method that exploits the second derivative of the RV (the jerk), which allows us to obtain a probability distribution for the companion's orbital period as well. In this section, all the quantities refer to the secondary companion, so we will avoid using sub-indices for clarity, except for the period $P_B$ and mass $M_B$ of the secondary. 

We begin with Equations (3) and (5) from \cite{rvimagingconstraints} to write the derivative of the radial velocity of the primary component $\dot{v}_{r}$ as
\begin{eqnarray}
\dot{v}_{r} &=& \frac{GM_B}{\ell^2} \Psi \label{eq:torres}\, ,\\
\Psi &=& [(1-e)(1+\cos E)]^{-1} (1-e \cos E) \sin i \times \nonumber \\
 & & (1-\sin^2(\nu+\omega)\sin^2 i) (1+\cos\nu)\sin(\nu+\omega) 
\,,
\end{eqnarray} 
where $\ell$ is the projected separation between the primary star and the companion in physical units (e.g., mks), and $\Psi$ is a rather intricate function that encapsulates all the orbital elements to be marginalized (time-dependencies included). Eq. \ref{eq:torres} can be evaluated by solving Kepler's equation
\begin{equation} 
\label{eq:kepler}
E-e\sin E = \mu 
\end{equation}
to obtain the eccentric anomaly $E$, and then using 
\begin{equation} 
\label{eq:nu}
\tan \frac{\nu}{2} = \sqrt{\frac{1+e}{1-e}}\tan \frac{E}{2} 
\end{equation} 
to derive the true anomaly $\nu$. To account for all possible combinations of periods and orbital phases, \cite{rvimagingconstraints} realized that the mean anomaly $\mu=\frac{2\pi}{P_B}t + \mu_{0}$ could be assumed to be uniformly distributed in $(0,2\pi]$. That is, irrespective of the values of the observation time $t$ and $P_B$, $\mu$ still can be assumed to have any orbital phase because $\mu_{0}$ can also have any value between $0$ and $2\pi$.
\begin{figure*}[t]
\centering
\subfloat[]{\label{fig:mvsp}\includegraphics[width=0.49\textwidth]{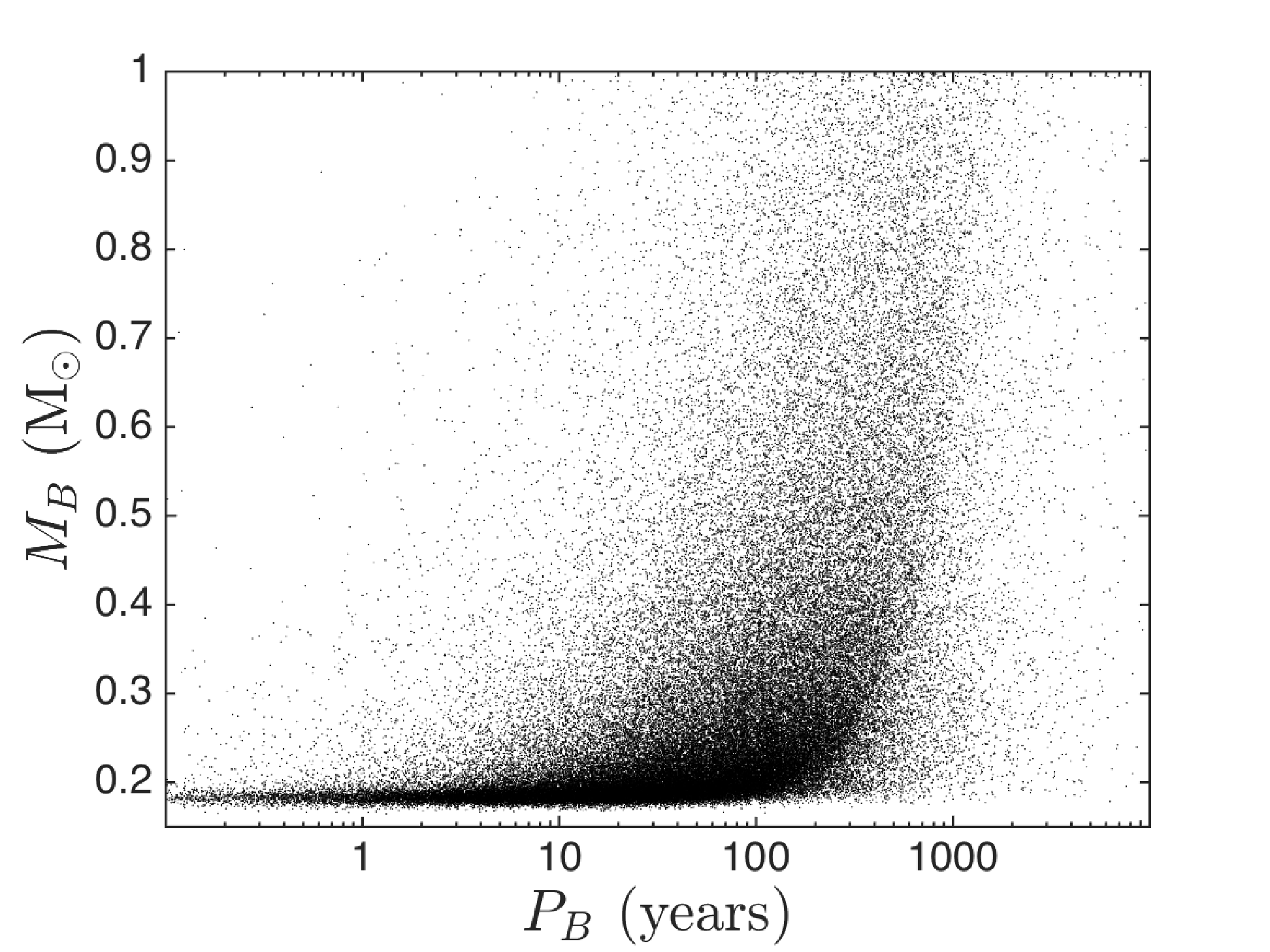}} 
\subfloat[]{\label{fig:m2}\includegraphics[width=0.49\textwidth]{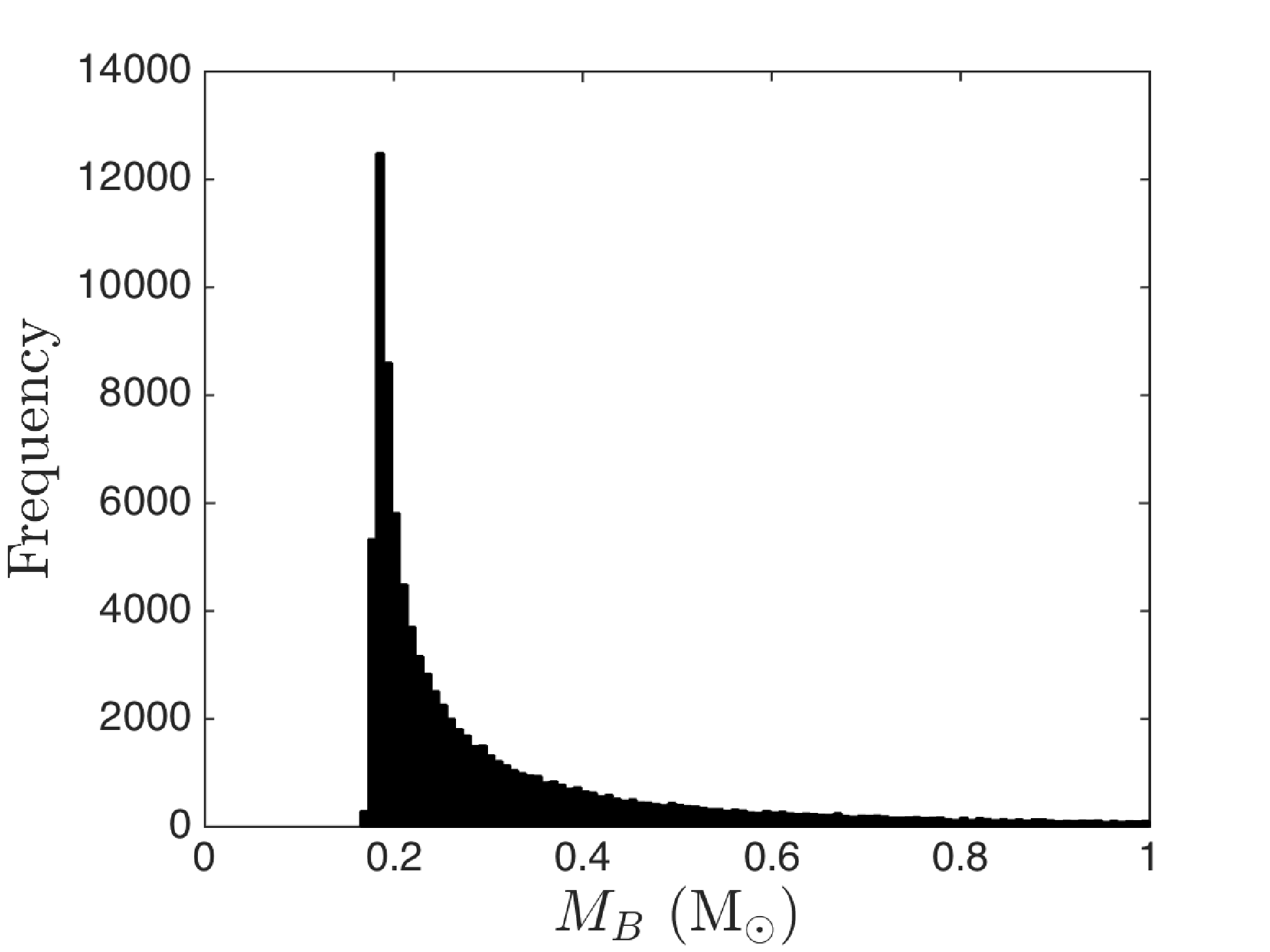}} \\
\subfloat[]{\label{fig:p2}\includegraphics[width=0.49\textwidth]{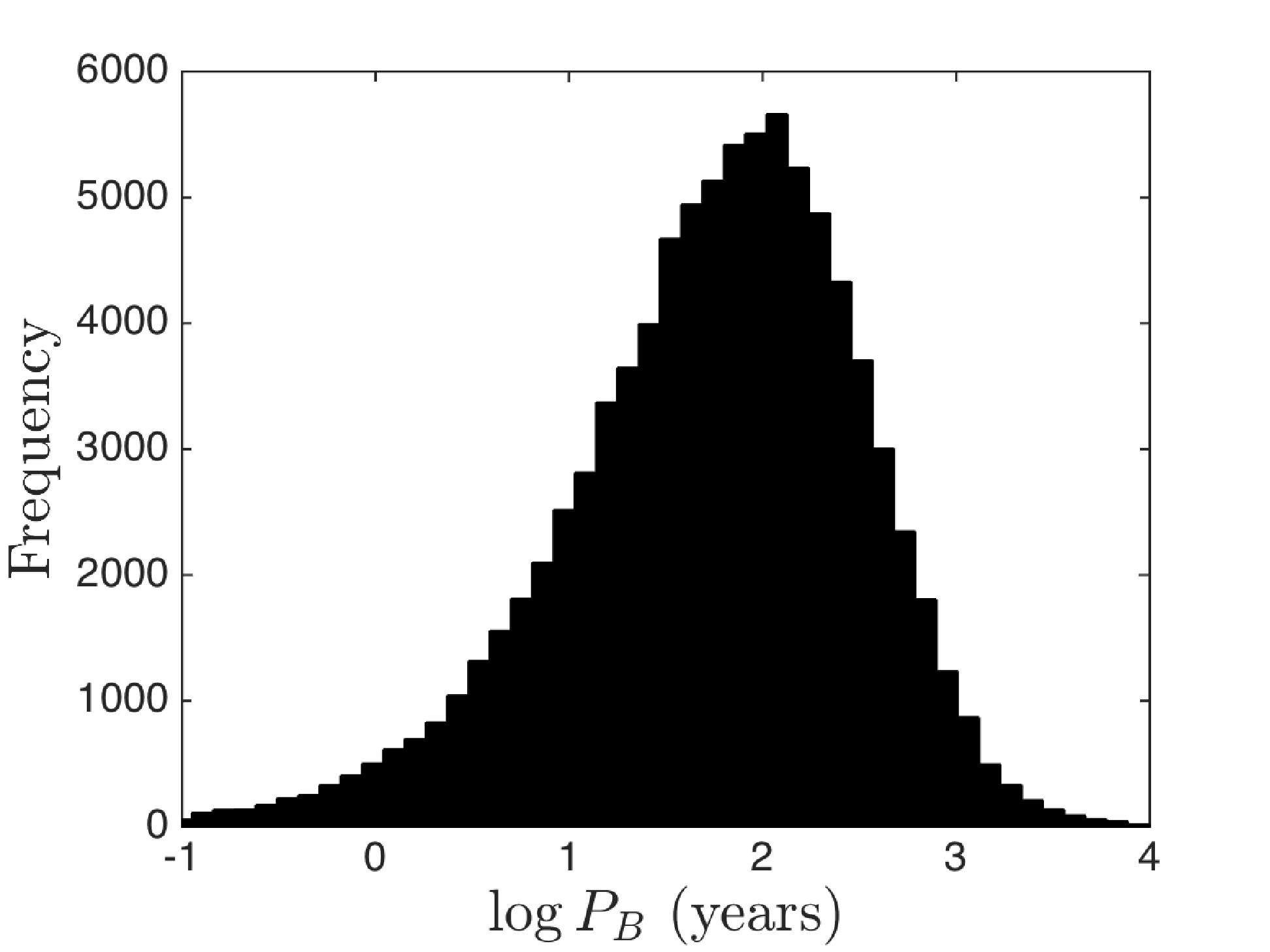}} 
\subfloat[]{\label{fig:i2}\includegraphics[width=0.49\textwidth]{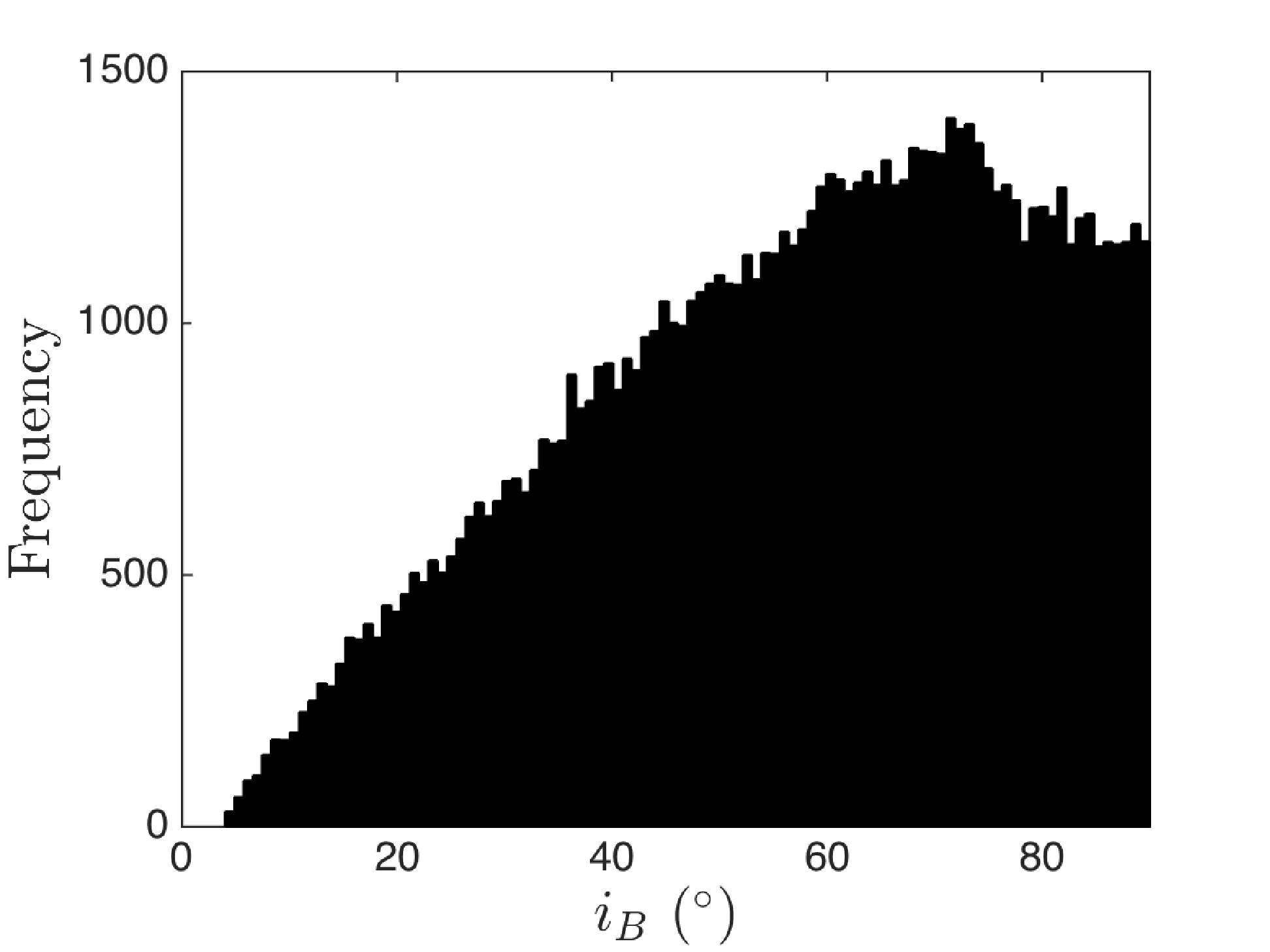}} 
\caption{(a) Mass vs. period for the outer companion computed using the MCMC method described in Section \ref{sec:stats}. (b) Mass distribution for the outer companion from the analysis, showing a sharp peak near \about 0.2 \msun. (c) $\log$ period distribution, showing a broad peak near \about 65 years. (d) Inclination distribution, showing a general preference for larger $i$.} 
\label{fig:dist2}
\end{figure*}

To use the information on the quadratic term in the RV curve, we first need to compute the second derivative of the RV. To do this efficiently, it is enough to realize that all the time dependence is included in $\mu$. Therefore, we can apply the chain rule and the fact that  $d\mu/dt = 2\pi/P_B$ to obtain
\begin{eqnarray}
\ddot{v}_r &=& 
2\pi\, \frac{G}{\ell^2}  \frac{M_B}{P_B} \Psi^\prime \label{eq:secondderivative} \\
\Psi^\prime &=& \frac{d\Psi}{d \mu} \,.
\end{eqnarray}
This is an important result because we have found that $\ddot{v}_r$ is proportional to $M_B/P_B$, and all the dependencies can again be marginalized by evaluating $\Psi^\prime$ (which is a function of time because it depends on $E$). While an analytic expression for $\Psi^\prime$ could be derived, it is far simpler (and requires fewer operations) to compute this numerically. We found that a simple two point formula with an infinitesimal increment of $10^{-4}$ radians for $\mu$ works to sufficient precision. By rearranging terms in Eq. \ref{eq:torres} and combining Eq. \ref{eq:torres} with Eq. \ref{eq:secondderivative}, we find 
\begin{eqnarray}
\frac{M_B}{M_\odot} &=&  5.341 \times 10^{-6} \,\, \dot{v}_r
\,\,\left(\frac{\rho}{\Pi}\right)^{2} \frac{1}{\Psi} \,\,,       \label{eq:ftorres}       \\
\frac{P_B}{yr}      &=&  2\pi\,\, \frac{\dot{v}_r}{\ddot{v}_r}
\,\,\frac{\Psi^\prime}{\Psi} \,\,,    \label{eq:final} 
\end{eqnarray}
where $\rho$ and $\Pi$ are the projected separation and parallax in arcseconds, respectively. Eq. \ref{eq:ftorres} was already derived in \cite{rvimagingconstraints}. The additional relation that we present here (Eq. \ref{eq:final}) can be used to constrain a companion's period using the same observables and marginalization method outlined in \cite{rvimagingconstraints}. The numerical factors in the equations come from the numerical substitution of the gravitational constant $G$, and the choice of units in \cite{rvimagingconstraints}, which assumes that $\dot{v}_r$ is in m s$^{-1}$ yr$^{-1}$, and $\ddot{v}_r$ is in m s$^{-1}$ yr$^{-2}$. Note that Eqs. \ref{eq:torres} and \ref{eq:final} assume that we can produce a Taylor expansion of the RV near the epoch of the direct image(s). The most straightforward way to impose this condition is to set the reference epoch in Eq. \ref{eq:rvmodel} to $t_0=t_{\rm image}$, thus deriving consistent MCMC samples for $\dot{v}_r$. This consideration is unnecessary when no curvature is detectable in the RV curve, as the first derivative of the RV is then independent of time.

With Eqs. \ref{eq:torres} and \ref{eq:final} in hand, we set out to constrain the mass and period of the outer companion. All the quantities have uncertainties, including $\dot{v}_r$, $\ddot{v}_r$, $\rho$, and $\Pi$. To account for these, we applied an additional refinement to the marginalization procedure of \cite{rvimagingconstraints}. That is, in addition to drawing random values for $e$, $\mu$, and $\sin{i_B}$, we also drew randomly-generated values of the observables: $\dot{v}_r$ and $\ddot{v}_r$ pairs were drawn from randomly selected states of the MCMC, and Gaussian distributions consistent with the error on parallax and the error on projected separation (Table \ref{tab:phot}) were used to generate plausible pairs of $\Pi$ and $\rho$, respectively. We know that the outer companion must be less massive than the primary (or it would be a known visual binary), so we excluded all parameters that corresponded to mass $>$ 1 \msun. Subsamples of the resulting distributions for the outer companion's mass, period, and inclination are shown in Fig. \ref{fig:dist2}. Based on these distributions, the median mass of HD 7449B is 0.23 \msun, and the mass of HD 7449B is larger than 0.17 \msun ~with a 99\% probability, consistent with our constraints from photometry. The median period is 65.7 years, corresponding to a semimajor axis of 17.9 AU. The inclination distribution is broad and has a median value of $i = 59.7$\degrees ~because it is mostly inherited from the uniform distribution in the Monte Carlo generated test values. However, because we exclude masses larger than 1 \msun, values of $i_B$ $<$ 8.4\degrees ~are also excluded with 99$\%$ probability, thus ruling out strictly face-on orbits. We do not show the distributions for $e$ and $\omega$ because they were completely unconstrained. Table \ref{tab:params} lists the outer companion's constrained parameters and ranges.


\subsection{Dynamical Constraints}
\label{sec:dynamics}
Based on the above analysis, we were able to constrain the outer companion's mass, period, and inclination, but not its eccentricity. To get a sense of the allowed ranges, we explored the dynamical stability of the system. We used the MERCURY integration package \citep{mercury} with a Bulirsch-Stoer integrator and simulated 100 different realizations of the system for 1 Gyr. The initial semimajor axis and eccentricity of the inner planet (HD 7449Ab) were held fixed at 2.32 AU, and 0.78, respectively, while its mass was set to 1 \mj\footnote{These values are slightly smaller than the nominal values listed in Table \ref{tab:params} to ensure that the limits on dynamical stability are conservative.}. The outer companion's semimajor axis was fixed at 18 AU and we assumed near-coplanarity such that its initial inclination relative to the planet was randomly drawn from values between 0 and 1\degrees \footnote{A very small initial inclination was chosen to avoid making the calculation completely 2D, which would preclude any possible inclination growth.}. In each of the 100 simulations, the companion's eccentricity was varied between 0 and 1 in increments of 0.01. Its mass was set to 0.17 \msun. For both the outer companion and the planet, the arguments of pericenter, longitudes of ascending node, and mean anomalies were all drawn randomly from a uniform distribution in each simulation. 
\begin{figure}[h]
\centering
\includegraphics[width=0.49\textwidth]{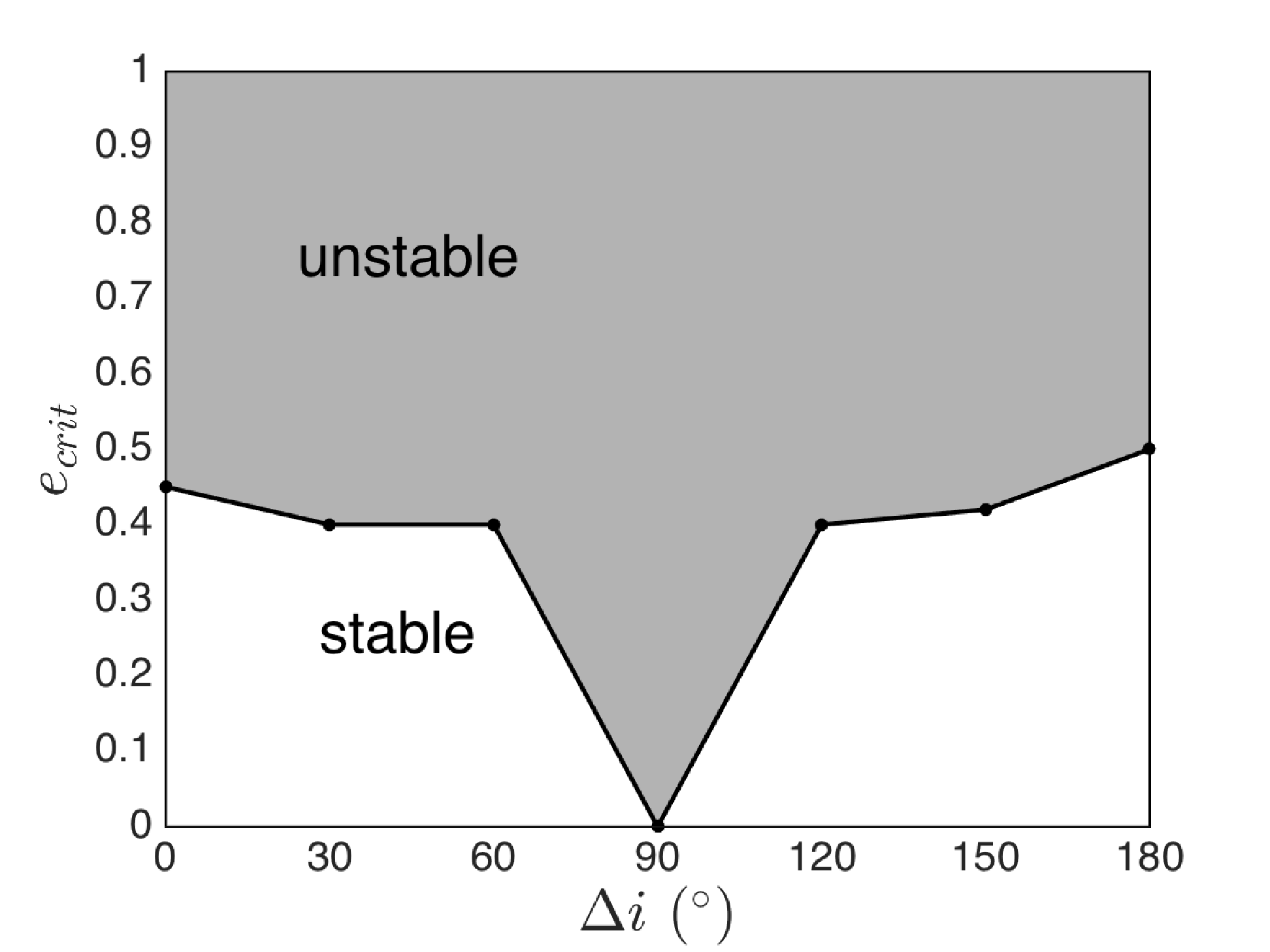}
\caption{Results of the numerical N-body simulations for HD 7449. The eccentricity of HD 7449B is constrained to be $\lesssim$ 0.5 for all mutual inclinations other than 90\degrees, for which the system is never stable.}
\label{fig:stability}
\end{figure}

In this case, the outer companion's critical eccentricity $e_{crit}$ (the eccentricity above which the planet's orbit becomes unstable) was 0.45. Based on this initial result, we performed additional simulations in which the planet and outer companion had mutual inclinations $\Delta i$ = 30$^{\circ}$, 60$^{\circ}$, 90$^{\circ}$, 120$^{\circ}$, 150$^{\circ}$, and 180$^{\circ}$. For the 90$^{\circ}$ case, the planet was never stable regardless of the outer companion's eccentricity. The critical eccentricities for the other inclinations were (in ascending order of mutual inclination) 0.4, 0.4, 0.4, 0.42, and 0.5, respectively. Based on these results, the outer companion's eccentricity is constrained to be $\lesssim 0.5$. Fig. \ref{fig:stability} summarizes the results of our stability analysis. 

\section{Discussion and Summary}
We have directly imaged the source of the long-period trend in the RV data for HD 7449. Based on our imaging, RV, and dynamical analysis, the outer companion HD 7449B is most likely a low-mass (\about 0.2 \msun) M dwarf orbiting at \about 18 AU with an eccentricity $\lesssim$ 0.5, although larger masses and periods cannot definitively be ruled out by the current data. We have also revised the parameters for the inner planet HD 7449Ab, finding that it is comparable in mass to Jupiter and on a very eccentric orbit. We find no evidence for additional planetary companions in the RV data. 

Now that HD 7449 is revealed to be a star-planet-M dwarf (SPM) binary, we can place it into relevant context. There are a handful of other SPM systems that consist of a planet orbiting one star with $a < 3$ AU and an M dwarf companion with $a$ \about 20 AU (e.g., HD 196885, \citealt{chauvinrvimaging}; $\gamma$ Cep, \citealt{gammacep}; Gliese 86, \citealt{gliese86}). HD 7449 is unique among these for two reasons: the secondary component has the lowest mass (\about 0.2 \msun ~compared to $> 0.4$ \msun ~for the others), and the inner planet is by far the most eccentric (0.8 compared to $< 0.5$ for the others). While core accretion is thought to be more difficult in systems like this, it should be possible to grow giant cores within \about 3 AU \citep{gammacepformation}. Furthermore HD 7449B's lower mass would be expected to cause less severe perturbations and thus have fewer detrimental effects on planet formation in the circumstellar disk. Perhaps this explains how the inner planet was able to form relatively unhindered.


How did the inner planet acquire such a large eccentricity? One possibility is the Kozai mechanism \citep{kozai, wumurray03}. If the planet and outer companion were initially on mutually-inclined orbits of at least 39.2\degrees, then the planet's eccentricity and inclination would oscillate with oppositely-occurring minima and maxima \citep{holmankozai}. Based on the nominal parameters for the planet and M dwarf companion, the length of a Kozai cycle would be \about a few hundred years, which is certainly short enough to be plausible given the age of the system (\about 2 Gyr). 

Assuming Kozai cycles are responsible, we can use the planet's current high eccentricity to constrain both the initial and current mutual inclination ($\Delta i_{init}$ and $\Delta i$). It can be shown that if the planet's orbit is initially circular, the maximum eccentricity is given by $e_{max} = \sqrt{1 - 5/3 \cos^{2}{\Delta i_{init}}}$ \citep{fabryckykozai}. For $e_{max}$ = 0.8, $\Delta i_{init}$ is constrained to be $\gtrsim$ 62\degrees. During Kozai cycles, the quantity $\sqrt{1 - e^{2}} \cos{i}$ of the planet is conserved. Using this relation, and the previous constraint on the initial mutual inclination, the current mutual inclination must be $\gtrsim$ 38\degrees. 

We can carry these constraints one step further. We know that the orbital inclination of the outer companion $i_B$ must be $>$ 8.4\degrees ~from our MCMC analysis (Section \ref{sec:stats}) and that the current mutual inclination $\Delta i$ $>$ 38\degrees ~if the planet was initially on a circular orbit and has been undergoing Kozai oscillations. Therefore, under these assumptions, $i_b$ must be $\gtrsim$ 46.4\degrees. Plugging this into $m\sin{i_b}$ = 1.09 \mj, the mass of HD 7449Ab would be $\lesssim$ 1.5 \mj, making the planet a true Jupiter analog.

Another explanation for the planet's large eccentricity is planet-planet scattering in the inner parts of the system (e.g., \citealt{rasio}). In this case, one or more planets may have been ejected from the system, leaving behind the eccentric HD 7449Ab. This scattering scenario would require both the surviving planet and the scattered planet to be relatively massive (7--10 \mj) and the eccentricity damping of the original circumstellar disk to be small \citep{moorheadscatter}. Given the ``smoking gun" (the nearby M dwarf companion), it seems more likely that Kozai cycles are responsible. 

The inner planet's high eccentricity and small perihelion distance (0.47 AU) raise the possibility of tidal circularization. However, its long period prevents it from circularizing on timescales shorter than \about 10$^{15}$ years \citep{tidalcirc}, meaning that it should continue to undergo Kozai oscillations for the foreseeable future. 

This interesting system should continue to be monitored by both RV and imaging. The latter technique, in particular, can provide additional constraints on HD 7449B's orbit, potentially leading to estimates of its dynamical mass \citep{crepptrends5}. Its eccentricity and inclination could also be further constrained, which could in turn help further constrain the inner planet's inclination. This would then allow for estimates of the inner planet's true mass, which is still a sparsely-measured parameter for exoplanets. 

High-resolution spectroscopy would help narrow down the effective temperature and spectral type of HD 7449B. While somewhat circular, this could be used to refine the photometry-derived mass (0.1-0.2 \msun), which then would affect the possible orbital configurations. For example, excluding RV solutions (from Section \ref{sec:stats}) that have mass $>$ 0.5 \msun ~leads to a median semimajor axis of \about 15 AU. Excluding masses $>$ 0.35 \msun ~corresponds to a median semimajor axis of \about 13 AU. Such small orbits would make HD 7449 a very tightly packed system with vigorous dynamical interactions and would require even more stringent constraints on the outer companion's eccentricity. Specifically, based on additional numerical N-body simulations we performed (using the same approach as described in Section \ref{sec:dynamics}), the eccentricity would have to be $\lesssim 0.3$ in these cases. 

Finally, the companion HD 7449B is interesting because it can become a benchmark object for future studies of stellar structure. The system represents a (still rare) case of an M dwarf with a measured age (via the primary) and a soon-to-be measured mass (via astrometric monitoring). The object's metallicity can be inferred from the primary's or could also be estimated using high-resolution spectroscopy. These quantities together can then help improve stellar structure models for similar cool stars (e.g., \citealt{baraffe15}), for which significant uncertainties still remain.

\acknowledgments 
We thank the anonymous referee for helpful comments and suggestions. We thank Andrew Tribick and Alexander Venner for notifying us of a unit conversion mistake prior to the article's publication. T.J.R. acknowledges support for Program number HST-HF2-51366.001-A, provided by NASA through a Hubble Fellowship grant from the Space Telescope Science Institute, which is operated by the Association of Universities for Research in Astronomy, Incorporated, under NASA contract NAS5-26555. Support for the D.M. is provided by the BASAL CATA Center for Astrophysics and Associated Technologies through grant PFB-06, by the Ministry for the Economy, Development, and Tourism's Programa Iniciativa Científica Milenio through grant IC120009, awarded to the Millennium Institute of Astrophysics (MAS), and by FONDECYT No. 1130196.

\bibliographystyle{apj}
\bibliography{ms}

\end{document}